\documentclass[12pt]{article}

\usepackage{amsmath,epsf}

\def\elabel#1{\label{#1}}

\newcommand{\be}{\begin{eqnarray}}
\newcommand{\ee}{\end{eqnarray}}
\newcommand{\nn}{\nonumber}
\newcommand{\ft}[2]{{\textstyle\frac{#1}{#2}}}
\newcommand{\eqn}[1]{(\ref{#1})}

\newcommand{\Bphi}{\boldsymbol{\phi}}
\newcommand{\Balpha}{\boldsymbol{\alpha}}

\newcommand{\Be}{\boldsymbol{e}}

\newcommand{\BN}{\boldsymbol{N}}
\newcommand{\Bh}{\boldsymbol{h}}
\newcommand{\BH}{\boldsymbol{H}}
\newcommand{\Bq}{\boldsymbol{q}}

\makeatletter
  \renewcommand{\theequation}{%
        \thesection.\arabic{equation}}
  \@addtoreset{equation}{section}
\makeatother

\begin{document}
\addtolength{\baselineskip}{2pt}

\begin{titlepage}

\begin{flushright}
SWAT-217 \\
UW/PT 99-04 \\
TIFR/TH/99-02 \\
hep-th/9902134
\end{flushright}
\begin{centering}
\vspace{.2in}
{\Large {\bf The BPS Spectra of Gauge Theories in \\ 
Two and Four Dimensions}}\\
\vspace{.4in}
Nicholas Dorey${}^{1\mbox{,}4}$, Timothy J. Hollowood${}^{2\mbox{,}4}$ and David Tong${}^{3}$ \\
\vspace{.3in}
$^{1}$Department of Physics, University of Washington, Box 351560   \\
Seattle, Washington 98195-1560, USA\\ 
{\tt dorey@phys.washington.edu} \\
\vspace{.2in}
$^2$Theoretical Division T-8, Los Alamos National Laboratory,\\
Los Alamos, NM 87545, USA\\
{\tt pyth@schwinger.lanl.gov}\\
\vspace{.2in}
$^3$Tata Institute of Fundamental Research, \\
Homi Bhabha Rd, Mumbai, 400 005, India\\
{\tt tong@theory.tifr.res.in} \\
\vspace{.2in}
$^4$Department of Physics, University of Wales, Swansea \\
Singleton Park, Swansea, SA2 8PP, UK\\
\vspace{.3in}
{\bf Abstract} \\
\end{centering}
We study ${\cal N}=(2,2)$ supersymmetric 
abelian gauge theories 
in two dimensions. The exact BPS spectrum of these models 
is shown to coincide with the spectrum of massive hypermultiplets 
of certain ${\cal N}=2$ supersymmetric gauge theories in four dimensions. 
A special case of these results involves a 
surprising
connection between four-dimensional ${\cal N}=2$ SQCD with $N$ colours and 
$N_{f}>N$ flavours
at the root of the baryonic Higgs branch and the supersymmetric 
$CP^{2N-N_{f}-1}$ $\sigma$-model in two dimensions. This correspondence 
implies a new prediction for the strong-coupling spectrum of the 
four-dimensional theory.


\end{titlepage}

\section{Introduction}
 
It has been noted many times in the past that two-dimensional theories 
with ${\cal N}=(2,2)$ supersymmetry have intriguing similarities to 
supersymmetric gauge theories in four dimensions. In this paper we will 
describe a quantitative correspondence between ${\cal N}=(2,2)$ theories 
in two dimensions and certain gauge theories 
with ${\cal N}=2$ supersymmetry in 
four dimensions which suggests new results for both types of model. 
In the present context, the key feature 
shared by these models is the existence of 
BPS states. The specific theories we will consider are, 
 
{\bf Theory A:} Two-dimensional ${\cal N}=(2,2)$ supersymmetric $U(1)_G$ 
gauge theory with dimensionful gauge coupling $e$. 
The matter content consists of $N$ 
chiral multiplets with charge $+1$ under $U(1)_G$ and 
{\em twisted} masses $m_1,\ldots,m_{N}$ \cite{hh}. 
A further $\tilde{N}<N$ 
chiral multiplets have charge $-1$ and twisted masses 
$\tilde{m}_{1},\ldots,\tilde{m}_{\tilde{N}}$. 
The classical theory has a dimensionless 
Fayet-Iliopoulis (FI) parameter $r$ and vacuum angle $\theta$ 
which are combined in a single complex coupling $\tau=ir+\theta/2\pi$. 
In the quantum theory, $\tau$ is eliminated in favour of a 
dynamical scale, $\Lambda$. 
 
{\bf Theory B:} Four-dimensional ${\cal N}=2$ $SU(N)$ SQCD with 
$N_f=N+\tilde{N}$ fundamental hypermultiplets with masses 
$m_1,\ldots,m_{N}$ and $\tilde{m}_1,\ldots,\tilde{m}_{\tilde{N}}$. The vacuum 
expectation value of the vector multiplet scalar is taken 
such that the theory lies at the root of its first baryonic 
Higgs branch. The dimensionless gauge coupling $g$ and vacuum 
angle $\vartheta$ combine to form the complex coupling 
$\tau=4\pi i/g^{2}+\vartheta/2\pi$ which is a parameter of the 
classical theory. In the quantum theory, $\tau$ is eliminated 
in favour of a dynamical scale, $\Lambda$.
 
The main result presented below is that the BPS spectrum of Theory A 
is identical to the spectrum of massive BPS-saturated hypermultiplets 
of Theory B. By this we mean that the masses of corresponding 
BPS multiplets in each theory are identical as functions of the 
parameters\footnote{As explained below and in \cite{nick} the BPS 
spectrum of Theory A does not depend on the two-dimensional 
gauge coupling $e$.} $m_{i}$, $\tilde{m}_{\tilde{i}}$ and $\Lambda$.  
A simplified version of the same correspondence also holds at the 
classical level with dependence on $\Lambda$ replaced by dependence 
on $\tau$. Similar results for the special case $\tilde{N}=0$ were 
presented in \cite{nick}. In this paper we extend the correspondence 
to the general case described above and investigate the consequences 
for the strong-coupling spectrum of the four-dimensional theory. 
We also discuss the correspondence of the BPS states in terms of the 
intersecting brane configuration introduced in \cite{hh} and its counterpart 
for the four-dimensional theory \cite{witm5}.   
     
The first step in demonstrating the claimed correspondence is to show 
that the central charges of both theories are identical as functions 
of the parameters. The central charge of the four-dimensional model 
is governed by the periods of an elliptic curve which is given explicitly in 
\cite{aps}.  In contrast, the two-dimensional 
central charge is determined by the critical values of a   
twisted superpotential. A key property of the two-dimensional 
BPS spectrum is its stability against D-term perturbations including, 
in particular, variation of the gauge coupling $e$ \cite{cfiv,CV2}. 
Because of this 
property, the exact twisted superpotential of Theory A and 
therefore its exact central charge can be determined by a one-loop 
calculation performed at weak coupling. We find that the central charge 
of both models is given by the same holomorphic function of $m_{i}$, 
$\tilde{m}_{\tilde{i}}$ and $\Lambda$. 
 
To check that the BPS spectra of the two theories agree, it is also 
necessary to specify which BPS states occur in each model. If we choose  
the parameters $m_{i}$ so that a sufficient number of gauge 
bosons have masses much greater than $|\Lambda|$, 
the four-dimensional theory is weakly 
coupled. In this case, the BPS spectrum, which can be determined by 
semiclassical methods, includes quarks, monopoles and dyons which are 
${\cal N}=2$ hypermultiplets in addition to the gauge bosons which lie 
in ${\cal N}=2$ gauge multiplets. We also find a characteristic 
spectrum of quark-monopole boundstates which have not been described before.  
In fact it is easy to show that Theory 
A, which is also weakly-coupled in this region of parameter space, contains 
BPS states corresponding to each massive quark, monopole and 
dyon hypermultiplet of Theory B, as well as counterparts of 
each of the quark-monopole boundstates mentioned above.  
 
Away from weak-coupling, the problem of determining which BPS states 
occur in the spectrum is complicated by the presence of curves of marginal 
stability on which BPS states may decay. For ${\cal N}=2$ theories in 
four dimensions, this problem has only been solved for gauge group 
$SU(2)$ \cite{bf}. However, in the two dimensional case, we can make some 
progress by noting that Theory A effectively reduces to the supersymmetric 
$CP^{N-\tilde{N}-1}$ $\sigma$-model at the special point in the 
strong-coupling region of parameter space 
where all twisted masses vanish\footnote{As explained in Section 3 below, 
if $\tilde{N}\neq 0$ Theory A also has a massless spectrum and 
it is necessary to introduce certain additional 
complex mass parameters to reduce the theory to the 
$\sigma$-model in question.}. This model is integrable and its exact 
spectrum is well known \cite{cpn,EH,CV2}. The $\sigma$-model has an unbroken 
$SU(N-\tilde{N})$ global symmetry and the BPS states form multiplets of 
this symmetry. In addition to kinks transforming in the fundamental 
representation, there is a characteristic  
spectrum of boundstates transforming in antisymmetric tensor 
representations of $SU(N-\tilde{N})$. Small twisted masses break the global 
symmetry and introduce mass splittings for these multiplets 
which are exactly determined by the one-loop twisted superpotential 
described above. 
As the point $m_{i}=\tilde{m}_{\tilde{i}}=0$ is not singular, this 
description of the spectrum should be valid in some open region of the 
parameter space containing this point. This strong coupling 
spectrum of kinks and their boundstates is obviously quite different 
from the weak coupling spectrum described above. For example the former 
spectrum contains only a finite number of BPS states while the latter 
is infinite. As usual these differences can be resolved by the 
presence of curves of marginal stability which disconnect the parameter 
space into inequivalent regions. 
   
As the exact central charges of Theory A and Theory B are identical, 
the same curves of marginal stability occur in both theories. Combining this 
with the fact that the spectra 
of the two theories already agree at weak coupling, it seems very plausible 
that the correspondence should hold throughout the parameter space. Indeed 
in the $N=2$ case, where four-dimensional results are available, this  
agreement can be checked explicitly \cite{nick}. 
To strengthen this conclusion in the general case, we review the 
realization of Theory A \cite{hh} and Theory B \cite{witm5}, 
on the world-volumes of 
intersecting M-theory branes.  In either theory, 
BPS states appear as M2 branes whose boundaries lie on M5-branes. 
In both cases, the relevant boundaries naturally correspond to one-cycles 
on the same Riemann surface.    
 
The equivalence of two-dimensional and four-dimensional spectra considered 
here has interesting consequences for both theories. As in 
\cite{nick}, we learn that the two-dimensional theory displays many 
of the phenomena which are characteristic of ${\cal N}=2$ theories in 
four-dimensions. These include non-trivial monodromies of the spectrum in the 
complex parameter space, curves of marginal stability as well as 
strong-coupling vacua with massless solitons. Conversely, 
setting all twisted masses to zero we uncover an unexpected relation 
between four-dimensional $SU(N)$ ${\cal N}=2$ SQCD with $N_{f}=N+\tilde{N}$ 
massless flavours at the baryonic root and the 
$CP^{N-\tilde{N}-1}$ $\sigma$-model in two dimensions. Both models are 
asymptotically free and have a $U(1)$ R-symmetry which is 
broken to a $Z_{2N-2\tilde{N}}$ subgroup by instantons\footnote{
Note that the correspondence discussed here is quite distinct from the more 
conventional analogy between the supersymmetric $CP^{N-1}$ $\sigma$-model 
in two dimensions and ${\cal N}=1$ 
supersymmetric Yang-Mills theory with gauge group $SU(N)$ in 
four dimensions}. The global $SU(N-\tilde{N})$ symmetry of the $\sigma$-model 
is identified with the unbroken global symmetry on the baryonic branch of the 
four-dimensional theory. The correspondence described above then makes  
a prediction for the four-dimensional BPS spectrum at the baryonic root. 
In addition to the massless spectrum at the baryonic root which was 
described in \cite{aps}, we should find massive 
BPS hypermultiplets transforming in the fundamental representation of the 
global $SU(N-\tilde{N})$ symmetry as well as 
boundstates in each anti-symmetric tensor representation of this group.  
Interestingly, an explicit semiclassical shows 
that these states are also present in the 
weak-coupling regime of the same four-dimensional theory\footnote{In other 
words, the theory with zero bare masses obtained from Theory B by moving away 
from the baryonic root to the region where the adjoint VEVs are 
much larger than $|\Lambda|$}.   
  
The plan of the paper is as follows. In section 2 we discuss various 
classical aspects of the two dimensional theory including 
a description of the twisted masses, the moduli space of classical vacua, 
the central charge and the 
spectrum of elementary quanta and solitons. In order 
to describe the latter in a manifestly supersymmetric manner, 
it is necessary to work in a dual Landau-Ginzburg 
formalism of the theory which we describe. In section 3, 
we turn to the quantum 
aspects of Theory A and determine the 
exact central charge and, in certain regimes 
of the parameter space, the BPS spectrum. In section 4, we 
show that the central charge of Theory A does indeed coincide 
with that of Theory B. We further describe the four-dimensional 
massive hypermultiplet spectrum in the weak-coupling 
regime and show that it agrees with the two-dimensional spectrum. 
Finally, in Section 5, we consider the IIA/M-brane description of the two 
theories.  

\section{Theory A: Classical Analysis}
 
Two-dimensional gauge theories with ${\cal N}=(2,2)$ supersymmetry were 
studied in detail by Witten in \cite{witn=2} and an extensive introduction to 
these theories, including the ${\cal N}=(2,2)$ superspace formalism used 
below, can be found in this 
reference. Additional relevant background material is given 
in \cite{hh,nick}. The conventions used are those of \cite{nick}.   
The two-dimensional gauge theory 
described in the introduction is built from  
a vector superfield $V$, 
$N$ chiral superfields $\Phi_i$, 
$i=1,\ldots,N$, with charge $+1$ under the gauge group $U(1)_G$, 
and $\tilde{N}$ 
chiral superfields $\tilde{\Phi}_{\tilde i}$, 
${\tilde{i}}=1,\ldots,\tilde{N}$, with charge $-1$. 
The gauge field kinetic terms are written most simply in terms 
of a gauge-invariant twisted chiral superfield, 
$\Sigma=D_+^\dagger D_-V$, whose lowest component is 
a complex scalar $\sigma$ and also includes the $U(1)$ field 
strength as well as fermionic superpartners. $\Sigma$ is referred to as 
the field-strength superfield of the corresponding gauge multiplet $V$.
Each chiral multiplet $\Phi_i$ ($\tilde\Phi_{\tilde{i}}$) 
consists of a complex scalar, $\phi_i$ 
($\tilde{\phi}_{\tilde{i}}$), and a single Dirac fermion. 
 
The kinetic terms for all fields are contained in an  
${\cal N}=(2,2)$ D-term,
\be
{\cal L}_{K}=\int {\rm d}^4\theta\ \left[ \sum_{i=1}^{N}
\Phi^\dagger_ie^{2V}\Phi_i+\sum_{\tilde{i}=1}^{\tilde{N}}
\tilde{\Phi}^{\dagger}_{\tilde{i}}e^{-2V}\tilde{\Phi}_{\tilde{i}}
-\frac{1}{4e^2}\Sigma^\dagger\Sigma\right]
\nn\ee
where $e$ is the dimensionful gauge coupling constant. Two 
dimensionless parameters will also prove important 
in the story: the Fayet-Iliopoulis (FI) parameter, $r$, and 
the vacuum angle, $\theta$. These are simply incorporated in 
the superfield formalism via the twisted F-term \cite{witn=2},  
\be
{\cal L}_{FI}=\frac{i\tau}{2}\int {\rm d}^2\vartheta\ \Sigma \ \ + \ \ 
{\mbox h.c.}
\elabel{tft}
\ee
where $\tau=ir+\theta/2\pi$ and ${\rm d}^2\vartheta$ is the measure 
over the `twisted' half of superspace. Both the D- and F-terms above are 
invariant under the full $U(1)_{A}\times U(1)_{R}$ automorphism group 
of the ${\cal N}=(2,2)$ supersymmetry algebra. The theory also has 
a global flavour symmetry $H=U(N)\times U(\tilde{N})/U(1)_{G}$. 
 
Gauge theories with ${\cal N}=(2,2)$ supersymmetry 
admit two different types of 
supersymmetric mass parameters for chiral multiplets. 
The first arises by dimensional reduction of the corresponding 
4D mass term and has the form of an ordinary superpotential (as opposed to 
a twisted superpotential), 
\begin{equation}
{\cal L}_{\hat{m}}= \int\, d^{2}\theta\ \hat{m}_{i\tilde j} \Phi_{i}
\tilde{\Phi}_{\tilde{j}} \,\,\, 
+ \,\,\, {\rm h.c.}
\elabel{complexm}
\end{equation} 
Following \cite{hh} we refer to the parameters $\hat{m}_{i\tilde{j}}$ 
as complex masses. Generic non-zero complex masses break $U(1)_{R}$ 
and also completely break the global symmetry group $H$. The R-symmetry 
$U(1)_{A}$ is left unbroken. The second kind of mass parameter, 
known as a twisted mass, 
has no counterpart in four dimensions. Twisted masses are introduced by 
weakly gauging the global flavour symmetry $H$ of the 
model and constraining the lowest component of the 
corresponding field-strength multiplet to a fixed 
background expectation value. In fact, in order to preserve 
supersymmetry this expectation value must be diagonalisable 
and we will only gauge the Cartan subalgebra $C$ of the flavour group $H$. We have $C=\left(\otimes_{i=1}^{N}U(1)_i\right)\times\left(
\otimes_{\tilde{i}=1}^{\tilde{N}}\tilde{U}(1)_{\tilde{i}}\right)/U(1)_{G}$, 
where $U(1)_i$ is defined so that 
the chiral multiplet $\Phi_j$ has charge $+\delta_{ij}$ while the 
$\tilde{\Phi}_{\tilde{j}}$ are neutral. Similarly, under 
$\tilde{U}(1)_{\tilde{i}}$, the $\Phi_j$ are neutral while 
the $\tilde{\Phi}_{\tilde{j}}$ carry charge 
$-\delta_{\tilde{i}\tilde{j}}$. Modding out by $U(1)_{G}$ is necessary 
because because the sum of the generators of the global 
$U(1)$ subgroups defined above generates the 
original gauge group. 
 
The vector multiplets corresponding to the newly gauged $U(1)$'s 
are denoted as $V_i$, $i=1,\ldots,N$ and 
$\tilde{V}_{\tilde{i}}$, $\tilde{i}=1,\ldots,\tilde{N}$ and the 
D-term Lagrangian for the kinetic terms becomes
\be
\tilde{\cal L}_{K}=\int {\rm d}^4\theta\ \left[ \sum_{i=1}^{N}
\Phi^\dagger_ie^{2(V+V_i)}\Phi_i+\sum_{\tilde{i}=1}^{\tilde{N}}
\tilde{\Phi}^{\dagger}_{\tilde{i}}e^{-2(V+\tilde{V}_{\tilde{i}})}
\tilde{\Phi}_{\tilde{i}}-
\frac{1}{4e^2}\Sigma^\dagger\Sigma\right]
\elabel{ltildek}
\ee
As each of their component fields is to be constrained, a kinetic 
term for the new gauge fields would be redundant. Instead 
we introduce a twisted chiral multiplet, 
$\Lambda_i$ ($\tilde{\Lambda}_{\tilde{i}}$), for each new gauge 
multiplet, $V_i$ ($\tilde{V}_{\tilde{i}}$), adding to the Lagrangian 
the term, 
\be
{\cal L}_{LM} = \frac{i}{2}\int {\rm d}^2\vartheta\ \left[
\sum_{i=1}^{N} 
\Lambda_i(\Sigma_i-m_i)+\sum_{\tilde{i}=1}^{\tilde{N}}
\tilde{\Lambda}_{\tilde{i}}(\tilde{\Sigma}_{\tilde{i}}-
\tilde{m}_{\tilde{i}})\right] \ \ + \ \ \mbox{ h.c.}
\elabel{llm}
\ee
where $\Sigma_i$ ($\tilde{\Sigma}_{\tilde{i}}$) is the field 
strength superfield for $V_i$ ($\tilde{V}_{\tilde{i}}$). These twisted chiral 
superfields act as Lagrange multipliers for the constraints 
$\Sigma_i=m_i$ and $\tilde{\Sigma}_{\tilde{i}}=\tilde{m}_{\tilde{i}}$. 
They also constrain the corresponding gauge fields 
to be pure gauge. Since these field 
strengths are themselves twisted chiral superfields, $m_i$ 
and $\tilde{m}_{\tilde{i}}$ are referred to as twisted masses. 
In addition to breaking $H$ 
down to its Cartan subalgebra, generic non-zero twisted masses 
also break $U(1)_{A}$. The residual R-symmetry group in the presence 
of twisted masses is $U(1)_{R}$. Note that the above 
procedure for introducing twisted masses only makes sense if the subgroup 
of the global symmetry group which we are gauging is unbroken. 
For this reason, if the chiral field $\Phi_{i}$ appears in a complex 
mass term such as (\ref{complexm}), 
it cannot also have a non-zero twisted mass\footnote{More precisely the 
twisted masses of the two chiral superfields appearing in (\ref{complexm}) 
are not independent parameters but obey a certain linear constraint.}. 
In this paper we will be primarily interested in the theory with non-zero 
twisted masses and complex masses are henceforth set to zero unless 
stated otherwise. The superspace Lagrangian for the model of 
interest is therefore 
${\cal L}=\tilde{\cal L}_{K}+{\cal L}_{FI}+{\cal L}_{LM}$.  Note 
also that the sum of all twisted masses may 
be changed by a shift in $\sigma$ and we use this freedom to set 
$\sum_{i=1}^{N}m_{i}=0$ while leaving the masses 
$\tilde{m}_{\tilde{i}}$ unconstrained.   
 
An alternative expression for this Lagrangian which will prove 
useful below 
is ${\cal L}={\cal L}_{D}+{\cal L}_{F}$ where the D-term is written as, 
\begin{equation}
{\cal L}_{D}=\int\,d^{4}\theta\, 
\sum_{i=1}^{N}\,\left[\exp\left(X_{i}+X^{\dagger}_{i}+2U_{i}\right)
-2{\cal R}_{i}U_{i}\right] + \sum_{\tilde{i}=1}^{\tilde{N}}\,
\left[\exp\left(\tilde{X}_{\tilde{i}}+\tilde{X}^{\dagger}_{\tilde{i}}-
2\tilde{U}_{\tilde{i}}\right)
-2\tilde{\cal R}_{\tilde{i}}\tilde{U}_{\tilde{i}}\right] 
-\frac{1}{4e^{2}}\Sigma^{\dagger}\Sigma
\elabel{d12}
\end{equation}
with $\Phi_{i}=\exp(X_{i})$ and 
$\tilde{\Phi}_{\tilde{i}}=\exp(\tilde{X}_{\tilde{i}})$. 
In the above expression we have also redefined the gauge superfields 
as $U_{i}=V+V_{i}$ and $\tilde{U}_{\tilde{i}}=V+\tilde{V}_{\tilde{i}}$ 
and split the Lagrange multiplier superfields into real and imaginary 
parts according to, 
\begin{eqnarray}
\Lambda_{i}=i{\cal R}_{i}+\frac{\Theta_{i}}{2\pi} & \qquad{} \qquad{} 
&  \tilde{\Lambda}_{i}=i\tilde{\cal R}_{\tilde{i}}+
\frac{\tilde{\Theta}_{\tilde{i}}}{2\pi}
\elabel{randi}
\end{eqnarray} 
Notice that each gauge field $U_{i}$ ($\tilde{U}_{\tilde{i}}$) 
has Fayet-Iliopoulis term where the FI coupling is the imaginary part, 
${\cal R}_{i}$ ($\tilde{\cal R}_{\tilde{i}}$), of the 
twisted chiral superfield 
$\Lambda_{i}$ ($\tilde{\Lambda}_{\tilde{i}}$). 
Note also that we have 
chosen to write these terms in the conventional way as D-terms depending on 
the vector superfields $U_{i}$ ($\tilde{U}_{\tilde{i}}$)  
rather than as twisted F-terms analogous to (\ref{tft}) above which involve 
the corresponding field-strength superfields. The residual twisted F-terms 
are conveniently written in terms of a twisted superpotential as 
\begin{equation}
{\cal L}_{F} = \int d^{2}\vartheta\, 
{\cal W}(\Lambda_{i},\tilde{\Lambda}_{\tilde{i}},\Sigma) \,+ \,  \int 
d^{2}\bar{\vartheta} \, \bar{\cal W}
({\Lambda}^{\dagger}_{i},\tilde{\Lambda}^{\dagger}_{\tilde{i}}, 
\Sigma^{\dagger})
\elabel{duall}
\end{equation}
with, 
\begin{equation}
{\cal W}=\frac{i}{2}\left[
\tau\Sigma - \sum_{i=1}^{N} 
\Lambda_i(m_i+\Sigma)-\sum_{\tilde{i}=1}^{\tilde{N}}
\tilde{\Lambda}_{\tilde{i}}(\tilde{m}_{\tilde{i}}+\Sigma)\right]
\elabel{tft2}
\end{equation}
 
At this point we 
have a model which contains both chiral and twisted chiral superfields. 
These two kinds of superfield have exactly the same content in terms 
of component fields and 
it is sometimes possible to eliminate a chiral superfield in favour 
of a twisted chiral superfield (or vice versa) using the two-dimensional 
duality transformation introduced by Rocek and Verlinde \cite{RV}. 
In the following we will use this transformation to derive a 
dual formulation in which the Lagrangian has a simple 
Landau-Ginzburg form. This is particularly 
straightforward for the present model 
as the first step in the duality transformation of \cite{RV} 
is to gauge a single global symmetry generator for each chiral superfield 
exactly as we have done above. Further in the usual duality transformation 
one must also introduce a Lagrange multiplier to enforce the constraint that 
the new gauge fields are pure gauge. This step also has already been 
performed above in the process of introducing twisted masses. The fact 
that the scalar components of these gauge multiplets are constrained to 
a non-zero value in the present case means this is a slight generalization 
of the usual duality transformation.   
 
To complete the duality transformation requires two more steps. First 
the chiral multiplets $X_{i}$ and $\tilde{X}_{\tilde{i}}$ are absorbed by the 
gauge transformation, 
\begin{eqnarray}
U_{i} & \rightarrow  & U_{i}-\frac{1}{2}(X_{i}+X^{\dagger}_{i}) \nonumber \\
 \tilde{U}_{\tilde{i}} & \rightarrow  & \tilde{U}_{\tilde{i}}
+ \frac{1}{2}(\tilde{X}_{\tilde{i}}+\tilde{X}^{\dagger}_{\tilde{i}})
\elabel{gtransmn}
\end{eqnarray}
Second we can eliminate the gauge multiplets $U_{i}$ and 
$\tilde{U}_{\tilde{i}}$ 
via their equations of motion which read, 
\begin{eqnarray}
\exp(2U_{i})={\cal R}_{i} & \qquad{} \qquad{} & 
\exp(-2\tilde{U}_{\tilde{i}})=-\tilde{\cal R}_{\tilde{i}}
\elabel{eom}
\end{eqnarray}
The result is a dual description of the theory in terms of the 
chiral superfields 
$\Lambda_{i}$, $\tilde{\Lambda}_{\tilde{i}}$
and $\Sigma$ with Lagrangian of Landau-Ginzburg form. We have 
${\cal L}={\cal L}_{D}+{\cal L}_{F}$ with,   
\begin{equation}
{\cal L}_{D}=\int d^{4}\theta \, 
{\cal K}[\Lambda_{i},\Lambda^{\dagger}_{i},\tilde{\Lambda}_{\tilde{i}},
\tilde{\Lambda}^{\dagger}_{\tilde{i}},\Sigma ,\Sigma^\dagger ] 
\elabel{dualld}
\end{equation}
where the K\"{a}hler potential is given by, 
\begin{equation}
{\cal K}=\sum_{i=1}^{N}\,{\cal R}_{i}\left(1-\log{\cal R}_{i}\right)
-\sum_{\tilde{i}=1}^{\tilde{N}}\,\tilde{\cal R}_{\tilde{i}}\left(1-
\log(-\tilde{\cal R}_{i})\right)-\frac{1}{4e^{2}}\Sigma^{\dagger}\Sigma
\elabel{dualk}
\end{equation}
The twisted F-term Lagrangian ${\cal L}_{F}$ is defined in 
(\ref{duall},\ref{tft2}) above. 
This dual formulation of the model will be particularly useful when 
discussing the properties of BPS solitons below. A limit of the theory 
which will be of particular interest below is the strong-coupling limit 
$e\rightarrow \infty$. In this limit the kinetic terms for the 
gauge multiplet can be omitted from the K\"{a}hler potential (\ref{dualk})  
and the field-strength multiplet $\Sigma$ 
can be eliminated via its equation of motion 
which imposes the linear constraint, 
\begin{equation}
\sum_{i=1}^{N} \Lambda_{i} + 
\sum_{\tilde{i}}^{\tilde{N}} \tilde{\Lambda}_{\tilde{i}}=\tau
\elabel{const}
\end{equation}
 
Returning to the original formulation of the model, with 
Lagrangian ${\cal L}=\tilde{\cal L}_{K}+
{\cal L}_{FI}+{\cal L}_{LM}$ defined above, 
we obtain the classical scalar potential by integrating out auxiliary 
fields to get, 
\be
U=e^2\left(\sum_{i=1}^{N}|\phi_i|^2-
\sum_{\tilde{i}=1}^{\tilde{N}}|\tilde{\phi}_{\tilde{i}}|^2-r\right)^2
+\sum_{i=1}^{N}|\sigma+m_i|^2|\phi_i|^2+\sum_{\tilde{i}=1}^{\tilde{N}}
|\sigma+\tilde{m}_{\tilde{i}}|^2|\tilde{\phi}_{\tilde{i}}|^2
\elabel{pot}\ee
The manifold of classical supersymmetric vacua, determined by 
the condition $U=0$, depends on the parameters $r$, $m_{i}$ and 
$\tilde{m}_{\tilde{i}}$. We begin by considering 
the case of zero twisted masses, $m_{i}=\tilde{m}_{\tilde{i}}=0$. 
In this case the theory has a classical Higgs branch with $\sigma=0$, 
determined by solving the equation,  
\be
\sum_{i=1}^{N}|\phi_i|^2-
\sum_{\tilde{i}=1}^{\tilde{N}}|\tilde{\phi}_{\tilde{i}}|^2=r
\elabel{eq1}
\ee
modulo $U(1)$ gauge transformations. The quotient of the solution space of 
(\ref{eq1}) by $U(1)_{G}$ 
defines a toric variety, ${\cal M}_{H}(r)$, 
of complex dimension $N+\tilde{N}-1$ \cite{witn=2,MP}. The generators of the 
global symmetry group H act as isometries on ${\cal M}_{H}$.
For $r>0$ ($r<0$) ${\cal M}_{H}$ has K\"{a}hler class $r$ ($-r$) 
and first Chern class $c_{1}=N-\tilde{N}$ ($\tilde{N}-N$). The two 
regions $r<0$ and $r>0$ are separated 
by the point $r=0$ at which the classical Higgs branch is singular. 
A special case is $N=\tilde{N}$, where the two regions yield a 
pair of birationally equivalent Calabi-Yau manifolds \cite{witn=2}. 
The classical Higgs branch is always non-compact except in the 
two cases, $\tilde{N}=0$ with $r>0$, and, $N=0$ with $r<0$, where the toric 
varieties in question are the complex projective spaces 
$CP^{N-1}$ and $CP^{\tilde{N}-1}$. 
Another special feature of the cases with $N=0$ ($\tilde{N}=0$), 
is that the classical theory has no supersymmetric vacua at all if $r$ 
is negative (positive). As we will see below, 
this and several other features of the classical theories 
are modified by quantum corrections. For $r=0$ only, 
the classical theory also has a Coulomb branch on which 
$\phi_{i}=\tilde{\phi}_{\tilde{i}}=0$ and $\sigma$ is unconstrained.   
 
When generic non-zero twisted masses are introduced the classical Higgs 
branch is lifted leaving only a finite number of isolated supersymmetric 
vacua. In particular, for $r>0$, we find $N$ such vacua, ${\cal V}_{i}$ 
with $i=1,2,\ldots, N$. In the vacuum ${\cal V}_{j}$ 
the scalar fields take values $\sigma=-m_{j}$, 
$\phi_i=\sqrt{r}\delta_{ij}$ and $\tilde{\phi}_{\tilde{i}}=0$. 
Correspondingly, for $r<0$ we find $\tilde{N}$ supersymmetric vacua, 
$\tilde{\cal V}_{\tilde{i}}$ 
with $\tilde{i}=1,2,\ldots, \tilde{N}$. In the vacuum 
$\tilde{\cal V}_{\tilde{j}}$ the scalar fields take values 
$\sigma=-\tilde{m}_{j}$, $\phi_{i}=0$, and 
$\tilde{\phi}_{\tilde{i}}=\sqrt{-r}\delta_{\tilde{i}\tilde{j}}$. 
In each of these vacua the scalar component of a single 
chiral multiplet is non-zero and the phase of this field has been set 
to zero by a $U(1)_{G}$ gauge rotation. When 
two or more twisted masses coincide a continuous vacuum degeneracy 
is restored. 
If $r>0$, the two minimal cases occur when $m_{i}=m_{j}$ for some $i$ and $j$ 
and when $m_{i}=\tilde{m}_{\tilde{j}}$ for some $i$ and $\tilde{j}$. Both 
cases give rise to Higgs branches of complex dimension one. In the first case  
the Higgs branch is a copy of $CP^{1}$ while in the second case it is a 
non-compact complex manifold which satisfies the Calabi-Yau condition 
$c_{1}=0$. In contrast, the condition 
$\tilde{m}_{\tilde{i}}=\tilde{m}_{\tilde{j}}$ for some $\tilde{i}$ and 
$\tilde{j}$ does not increase the vacuum degeneracy (for $r>0$).  
As in the massless case, the theory with $r=0$ also has a Coulomb 
branch on which all the chiral 
multiplet scalars vanish and $\sigma$ is unconstrained.
 
We will now consider the BPS spectrum of the classical theory introduced 
above beginning with the case of vanishing twisted masses. In this case 
the full R-symmetry group $U(1)_{A}\times U(1)_{R}$ is unbroken. This 
implies that the ${\cal N}=(2,2)$ SUSY algebra has vanishing central charge 
and thus there are no massive BPS states in the classical theory. However, 
the classical theory does have massless particles corresponding 
to the flat directions of the Higgs and Coulomb branches described 
above. In particular there are $N+\tilde{N}-1$ 
massless chiral multiplets which correspond to 
complex coordinates on the classical Higgs branch ${\cal M}_{H}(r)$. 
In any open region of 
${\cal M}_{H}$ in which $\Phi_{j}\neq 0$, the gauge-invariant superfields 
$W_{i}^{(j)}=\Phi_i/\Phi_j$, for $i=1,2,\ldots, N$ with $i \neq j$, 
together with $\tilde{W}^{(j)}_{\tilde{i}}= \Phi_{j}\tilde{\Phi}_{\tilde{i}}$, for $\tilde{i}=1,2,\ldots,N$, provide a convenient basis for these massless 
fields. In the classical theory 
the scalar components of some of these multiplets are Goldstone bosons for the 
broken generators of the global symmetry group H, a situation which 
cannot persist in 
the corresponding two-dimensional quantum theory \cite{Col2}. 
In contrast, fluctuations of the fields 
which are orthogonal to the vacuum manifold 
get masses of order $\sqrt{e}|r|$ from the 
Higgs mechanism. Provided $r\neq 0$, the massive 
fields decouple in the infra-red (IR) limit 
$e\rightarrow \infty$ and the resulting theory 
is precisely a supersymmetric non-linear $\sigma$-model with 
target manifold ${\cal M}_{H}(r)$. The coupling constant of the low-energy 
$\sigma$-model is related to the FI parameter as $g=\sqrt{2/r}$. 
 
In the presence of twisted 
masses, the same IR limit yields a massive deformation of the $\sigma$-model 
which was studied in detail for the case $\tilde{N}=0$ where 
${\cal M}_{H}=CP^{N-1}$ in \cite{nick}. In the deformed model, 
the classical vacuum degeneracy is lifted as described above 
and each of the $N+\tilde{N}-1$ massless chiral multiplets of the 
$\sigma$-model acquires a mass. As usual, the 
conversion of massless degrees of freedom to 
massive is only consistent if the latter states are BPS saturated and 
therefore lie in short multiplets 
of supersymmetry. This is possible due to the fact that twisted masses 
break $U(1)_{A}$ allowing a non-zero central charge in the supersymmetry 
algebra. Further the twisted masses also break $H$ to its maximal torus and 
the unbroken Cartan generators are natural candidates for the 
central charges in question \cite{hh}. 
Massive particles which carry the global $U(1)$ charges can 
then form short multiplets. In the vacuum 
${\cal V}_{j}$, the $N+\tilde{N}-1$ massless 
chiral multiplets $W_{i}^{(j)}$ and 
$\tilde{W}^{(j)}_{\tilde{i}}$ defined above get masses $|m_{i}-m_{j}|$ 
and $|\tilde{m}_{\tilde{i}}-m_{j}|$ respectively. Summing over each of 
the vacua ${\cal V}_{i}$ of the $r>0$ theory, we find a total of 
$N(N+\tilde{N}-1)$ states with distinct masses. Introducing a little 
new notation allows these masses to be 
written in a simple universal form. We denote the charge 
carried by a field under the global $U(1)_k$ 
($\tilde{U}(1)_{\tilde{k}}$) flavour symmetries 
as $S_{k}$ ($\tilde{S}_{\tilde{k}}$) 
Note that $W^{(j)}_{i}=\Phi_i/\Phi_j$ then 
carries charges $S_k=\delta_{ik}-\delta_{jk}$ and 
$\tilde{S}_{\tilde{k}}=0$, while the field  
$\tilde{W}_{\tilde{i}}^{(j)}=\tilde{\Phi}_{\tilde{i}}\Phi_j$ 
carries charges $S_k=\delta_{jk}$ and 
$\tilde{S}_{\tilde{k}}=-\delta_{\tilde{i}\tilde{k}}$. 
The masses of all these states obey the BPS mass formula 
$M=|Z_{S}|$ where the corresponding central charge is, 
\begin{equation}
Z_{S}= i\sum_{i=1}^{N}m_iS_i +i\sum_{\tilde{i}=1}^{\tilde{N}}\tilde{m}_
{\tilde{i}}
\tilde{S}_{\tilde{i}}
\elabel{cc1}
\end{equation}
 
The BPS states described above are elementary quanta of the fields appearing 
in the Lagrangian of the mass-deformed $\sigma$-model. As the classical 
theory with non-zero twisted masses has isolated supersymmetric vacua, 
an additional possibility arises: there can be 
BPS saturated kinks which interpolate 
between distinct supersymmetric vacua at left and right spatial infinity. 
For $r>0$, we can define topological charges $T_i$, $i=1,\ldots,N$, 
such that a field configuration that tends asymptotically 
to the vacuum ${\cal V}_j$ as $x\rightarrow\infty$ and to 
${\cal V}_k$ as $x\rightarrow -\infty$ has topological charge 
$T_i=\delta_{ij}-\delta_{ik}$. Correspondingly, for $r<0$, we define 
topological charges, $\tilde{T}_{\tilde{i}}$, for 
$\tilde{i}=1,2,\ldots,\tilde{N}$, associated with solitons 
which interpolate between the vacua $\tilde{\cal V}_{\tilde{i}}$.   
Like the global generators $S_{i}$ and 
$\tilde{S}_{\tilde{i}}$ defined above, the topological charges 
$T_{i}$ and $\tilde{T}_{\tilde{i}}$, can also contribute to 
the central charge \cite{WO}. Solitons with non-zero topological charges 
can then give rise to additional BPS states in the spectrum. 
As we review below, the topological contribution to the classical 
central charge (\ref{cc1}) is included by the replacement $S_{i}
\rightarrow S_{i}+\tau T_{i}$ in (\ref{cc1}) for $r>0$  
and $\tilde{S}_{\tilde{i}}
\rightarrow \tilde{S}_{\tilde{i}}+\tau\tilde{T}_{\tilde{i}}$ for $r<0$. 
 
For the case of the mass-deformed $CP^{N-1}$ $\sigma$-model, 
BPS saturated kinks were studied in detail in \cite{nick}. 
We will begin 
by briefly reviewing the simplest example, $N=2$, $\tilde{N}=0$, 
where the target space is $CP^{1}$. In this case there are two classical 
vacua ${\cal V}_{1}$ and ${\cal V}_{2}$, with a single  
chiral multiplet of mass $|m|=|m_{1}-m_{2}|$ in each. The 
scalar components of these multiplets are 
$w=\phi_{1}/\phi_{2}$ in ${\cal V}_{1}$ and $1/w$ in ${\cal V}_{2}$. 
The theory has a single global $U(1)$ charge $S=(S_{1}-S_{2})/2$ 
and a single topological charge $T=(T_{1}-T_{2})/2$.   
It will be useful to exhibit the soliton solutions in both versions of the 
model introduced above. Starting from the original 
formulation of the theory, 
with Lagrangian 
${\cal L}=\tilde{\cal L}_{K}+ {\cal L}_{FI}+{\cal L}_{LM}$ 
defined in (\ref{tft},\ref{ltildek},\ref{llm}), 
we take the IR limit $e\rightarrow \infty$ and eliminate 
the gauge field multiplet by its equations of motion. We then obtain a 
Lagrangian for the complex scalar $w=\phi_{1}/\phi_{2}$ and 
its superpartner. It is convenient to  
decompose the complex field $w$ in terms of its modulus and argument as, 
\begin{equation}
w=\tan\frac{\varphi}{2}\exp(i\alpha)
\elabel{cov2}
\end{equation}
where, in order to make the mapping one-to-one, we make the identifications 
$\varphi\sim\varphi+2\pi$ and $\alpha\sim\alpha+2\pi$.   
In terms of the new variables, the bosonic terms in the Lagrangian read 
\cite{nick},  
\begin{equation}
{\cal L}_{\rm Bose}= -\frac{r}{4}\left[(\partial_{\mu}\varphi)^{2}+
\sin^{2}\varphi\left(|m|^{2}-(\partial_{\mu}\alpha)^{2}\right)\right] +
\frac{\theta}{4\pi}\epsilon^{\mu\nu}\partial_{\mu}(\cos\varphi)\partial_{\nu}\alpha
\elabel{sg}
\end{equation}
This is a variant of the sine-Gordon (SG) Lagrangian with 
an additional massless field $\alpha$ which has 
derivative couplings to the SG field $\varphi$. The 
two SUSY vacua found above correspond to the two sets of zeros of the 
SG potential, which occur at $\varphi =2n\pi$ and at $\varphi=(2n+1)\pi$ 
for integer $n$. As $\alpha$ appears only through its derivatives it can 
take any constant value in the vacuum.       
 
The classical equations of motion coming from (\ref{sg}), have a 
family of solutions of topological charge $T=1$, 
with $\alpha=\omega t$ and, 
\begin{equation} 
\varphi= 2 \tan^{-1}\left(\exp\sqrt{|m|^{2}-\omega^{2}} \right)
\elabel{dyon}
\end{equation}         
which are labelled by the real parameter $\omega$ with $|\omega|<|m|$. 
The mass and global charge of the solution are, 
\begin{eqnarray}
M= \frac{r|m|^{2}}{\sqrt{|m|^{2}-\omega^{2}}}    
& \qquad{} \qquad{} \qquad{} &  S=\frac{r\omega}{\sqrt{|m|^{2}-\omega^{2}}}
\,-\, \frac{\theta}{2\pi}    
\elabel{mns}
\end{eqnarray}  
Applying semiclassical Bohr-Sommerfeld quantization to these periodic 
classical solutions we find that the allowed values of $\omega$ are 
quantized so that $S$ takes only integer values \cite{nick}. 
Hence the theory has an infinite tower of `dyons' which carry both 
topological charges and Noether charges.   
The contribution of the vacuum angle $\theta$ to the global charge 
is a two-dimensional analog of the Witten effect \cite{WE} for dyons in 
four dimensions. 
Eliminating the variable $\omega$ we find that the soliton mass 
can be written as 
\begin{equation}
M=|m|\sqrt{\left(S+\frac{\theta}{2\pi}\right)^{2}+r^{2}}
\elabel{sat2}
\end{equation}
>From the above we learn that the masses of both the dyons 
and the elementary quanta are consistent 
with the BPS mass formula $M=|Z|$ where, 
\begin{equation}
Z=im\left(S+\tau T \right)
\elabel{ccharge4}
\end{equation}
 
The description of the solitons given above and in \cite{nick} 
has a serious drawback: supersymmetry is not manifest and properties 
such as BPS saturation of the solutions 
have to be checked by explicit calculation. In particular, we would 
like to derive the formula (\ref{ccharge4}) for the central charge directly, 
rather than guessing it from the mass spectrum as we did above. 
In fact for a large class of ${\cal N}=(2,2)$ models with Lagrangians 
of Landau-Ginzburg form, the Bogomol'nyi bound can be derived directly 
in superspace \cite{FMVW}. 
As we showed above, the two-dimensional theory 
with twisted masses can be be put in 
this standard form by performing a Rocek-Verlinde 
duality transformation. In the case 
$N=2$, $\tilde{N}=0$, after imposing the constraint (\ref{const}), the  
dual Lagrangian can be written in terms of a single twisted 
chiral superfield, 
by setting $\Lambda_{1}=\Lambda$ and $\Lambda_{2}=\tau-\Lambda$. 
The resulting Lagrangian is
\begin{equation}
{\cal L}=\int d^{4}\theta \, 
{\cal K}[\Lambda,\Lambda^{\dagger}] \,+\, 
\int d^{2}\vartheta\,{\cal W}(\Lambda) \,+ \,  \int 
d^{2}\bar{\vartheta} \, \bar{\cal W}(\Lambda^{\dagger})
\elabel{lglag}
\end{equation}
with K\"{a}hler potential.     
\begin{equation}
{\cal K}[\Lambda]={\cal R}\left(1-\log{\cal R}\right)+
(r-{\cal R})\left(1-\log(r-{\cal R})\right)
\elabel{kpot}
\end{equation}
where $\Lambda=i{\cal R}+\Theta/2\pi$. In the following, $\Lambda$, 
${\cal R}$ and $\Theta$ will denote either the superfield or its scalar 
component depending on context. The K\"{a}hler metric is,  
\begin{equation} 
g_{\Lambda\bar{\Lambda}}=-\frac{\partial^{2} \cal K}
{\partial\Lambda\partial \Lambda^{\dagger}} =\frac{r}{4{\cal R}(r-{\cal R})}
\elabel{kmetric}
\end{equation}
and the twisted superpotential ${\cal W}$ has the 
simple form ${\cal W}=im\Lambda/2$. The Lagrangian (\ref{lglag}) is in fact 
the most general Lagrangian for a single twisted chiral superfield 
with at most two derivatives and we will meet it again with different 
choices for ${\cal K}$ and ${\cal W}$ in the next section. For this reason we 
will now give a general derivation of the Bogomol'nyi bound 
at the same time as giving the specific formulae which relate to the 
present choice of ${\cal K}$ and ${\cal W}$.      
 
After integrating out auxiliary fields the bosonic part of the 
above Lagrangian becomes, 
\begin{eqnarray}
{\cal L}_{\rm Bose} & = & -g_{\Lambda\bar{\Lambda}}
\partial_{\mu}\Lambda\partial^{\mu}\Lambda^{\dagger}-
g^{\Lambda\bar{\Lambda}}
\frac{\partial{\cal W}}{\partial\Lambda}
\frac{\partial\bar{{\cal W}}}{\partial\Lambda^{\dagger}} \nonumber \\
& = & -\frac{r}{4{\cal R}(r-{\cal R})}
\left[(\partial_{\mu}{\cal R})^{2}+
\frac{1}{4\pi^{2}}
(\partial_{\mu}\Theta)^{2}\right]-\frac{|m|^{2}}{r}{\cal R}(r-{\cal R}) 
\elabel{N=2lag}
\end{eqnarray}
where $g^{\Lambda\bar{\Lambda}}=1/g_{\Lambda\bar{\Lambda}}$. 
We can immediately compare this with the Lagrangian (\ref{sg}) 
for the bosonic fields $\varphi$ and $\alpha$ and deduce the 
identifications,  
\begin{eqnarray}
{\cal R} = r\sin^{2}\frac{\varphi}{2} & \qquad{} \qquad{} & 
\partial_{\mu}\Theta=\pi r\sin^{2}\varphi
\varepsilon_{\mu\nu}\partial^{\nu}\alpha
\elabel{ident}
\end{eqnarray} 
The second equality states that the scalar `field strengths' 
$\partial_{\mu}\Theta$ and $\partial_{\mu}\alpha$ 
are related by a two-dimensional analog of electric-magnetic duality 
which interchanges equations of motion and Bianchi identities 
\cite{RV}. The bosonic potential, 
\begin{equation} 
U=g^{\Lambda\bar{\Lambda}}
\frac{\partial{\cal W}}{\partial\Lambda}
\frac{\partial\bar{\cal W}}{\partial\Lambda^{\dagger}}=
\frac{|m|^{2}}{r}{\cal R}(r-{\cal R})    
\elabel{svac}
\end{equation}
has zeros at ${\cal R}=0$ and ${\cal R}=r$ corresponding to the two 
supersymmetric vacua of the model. The field $\Theta$ can take any 
constant value in the vacuum as it only its derivatives appear 
in the Lagrangian. Note that the zeros of $U$ come from poles 
in the K\"{a}hler metric $g_{\Lambda\bar{\Lambda}}$ rather than zeros of 
$\partial {\cal W}/\partial \Lambda$ which is more conventional. 
This does not indicate any pathology of the underlying model 
but rather reflects our choice of coordinates in field space.  
 
We will now consider solitons  which interpolate 
between between distinct 
vacua at left and right spatial infinity. Specifically we consider 
boundary conditions, $\Lambda\rightarrow 0$ as $x\rightarrow -\infty$ and 
$\Lambda\rightarrow \Lambda_{+}=ir+\Delta\Theta/2\pi$ 
as $x\rightarrow +\infty$. The mass of such a 
configuration obeys the following inequality \cite{FMVW} 
which hold for any complex constant $\gamma$ with $|\gamma|=1$, 
\begin{eqnarray}
M &= & \int_{-\infty}^{+\infty}\,dx\, \left[ g_{\Lambda\bar{\Lambda}}
\frac{\partial\Lambda}{\partial x }
\frac{\partial\Lambda^{\dagger}}{\partial x} + g^{\Lambda\bar{\Lambda}}
\frac{\partial{\cal W}}{\partial\Lambda}
\frac{\partial\bar{{\cal W}}}{\partial\Lambda^{\dagger}} \right] \nonumber \\
& = & \int_{-\infty}^{+\infty}\,dx\,\left|\frac{\partial\Lambda}
{\partial x }-\gamma g^{\Lambda\Lambda^{\dagger}}\frac{\partial\bar{{\cal W}}}
{\partial\Lambda^{\dagger}}\right|^{2} + 
\int_{-\infty}^{+\infty}\,dx\,\left[\bar{\gamma}
\frac{\partial{\cal W}}{\partial\Lambda}
\frac{\partial\Lambda}{\partial x }+\gamma
\frac{\partial\bar{{\cal W}}}{\partial\Lambda^{\dagger}}
\frac{\partial\Lambda^{\dagger}}{\partial x }\right] \nonumber \\ 
&\geq & \qquad{}  2{\rm Re}
\left[\bar{\gamma}\left({\cal W}(\Lambda_{+})-{\cal W}(0)\right)\right]
\elabel{bog1} 
\end{eqnarray}
By choosing $\gamma=\Delta{\cal W}/|\Delta{\cal W}|$ with $\Delta{\cal W}=
{\cal W}(\Lambda_{+})-{\cal W}(0)$ we obtain the Bogomol'nyi bound 
$M\geq 2|\Delta{\cal W}|$. This corresponds to a non-zero 
value for the central charge $Z=2\Delta{\cal W}=
im\Delta\Lambda=-mr+im\Delta\Theta/2\pi$. 
 
To interpret this formula recall that in the 
$(\theta,\alpha)$ variables we found time-dependent solutions (\ref{dyon}) 
with $\alpha=\omega t$. The identification (\ref{ident}) implies that 
time-dependence of $\alpha$ translates to $x$ dependence for $\Theta$.  
An explicit comparison of soliton solutions in the 
two sets of variables $(\varphi,\alpha)$ and $({\cal R},\Theta)$ is 
performed in Appendix A, where it is shown that the time-dependent 
dyon solutions (\ref{dyon}) in the first set of variables 
correspond to static BPS configurations which saturate the bound 
(\ref{bog1}) in the second. In particular the relevant 
boundary conditions for $\Theta$ are 
related to the global $U(1)$ charge $S$ as 
$\Delta\Theta =\theta+2\pi S$. 
The resulting BPS mass formula is $M=|Z|$ with 
central charge $Z=2\Delta{\cal W}=im (S+\tau)$ which 
agrees with (\ref{ccharge4}) 
in the sector of topological charge $T=1$. Note that the fact that 
$S$ is quantized in integer units, which is derived from the 
Bohr-Sommerfeld quantization condition in the $(\varphi,\alpha)$ 
variables, is not at all obvious in the $({\cal R},\Theta)$ variables. 
The two descriptions are analogous to the two different choices of gauge, 
due to Tomboulis and Woo (TW) \cite{TW} and to Julia and Zee (JZ) \cite{JZ} 
respectively, used to describe BPS 
dyon solutions in four-dimensions. In the TW gauge the dyon solution has 
periodic time-dependence, while in the JZ gauge it is static. 
Similarly, the quantization of electric charge is obvious in the 
first gauge but not in the second.   
 
As emphasised above and in \cite{nick}, 
the BPS spectrum of the mass-deformed supersymmetric 
$CP^{1}$ $\sigma$-model has strong similarities to that of 
a gauge theory with extended supersymmetry in four-dimensions. This 
correspondence becomes 
particularly clear in the $({\cal R},\Theta)$ variables introduced above. 
To exhibit this connection it is useful to recall another general 
property of  
${\cal N}=(2,2)$ solitons in theories of Landau-Ginzburg type. 
As shown in \cite{CV2}, the soliton always follows a particular 
trajectory in field space: a straight line in the complex ${\cal W}$ plane 
which joins two vacua. In the present case we have 
${\cal W}=im\Lambda/2$, so this is equivalent to a straight line-segment 
in the complex $\Lambda$-plane. This is demonstrated explicitly in 
Appendix A. 
The mass of the resulting BPS state is proportional 
to the length of this line-segment. As above we set $\Lambda=i{\cal R}+ 
\Theta/2\pi$ and note that vacua occur on the two lines ${\cal R}=0$ 
and ${\cal R}=r$. Further, it follows from the discussion above that 
the vacuum values of $\Theta$ are also quantized in units of of $2\pi S$.  
They can be chosen to lie at $\Theta=2\pi n$ for ${\cal R}=0$ and 
$\Theta=2\pi n+\theta$ for ${\cal R}=r$. We now have two infinite 
rows of vacua in the complex $\Lambda$-plane and each BPS state, including 
both elementary particles and dyons as well as the corresponding 
charge-conjugate states, is associated with a line segment joining two 
vacua. The mass of each state is the length of the line segment. 
This is essentially identical to the classical charge lattice of 
BPS states of a four-dimensional ${\cal N}=2$ theory\footnote{Recall that only BPS states of magnetic charge 
$0,\pm 1$ occur in $SU(2)$ 
${\cal N}=2$ supersymmetric Yang-Mills theory. Each of these states 
corresponds to a vector joining two vacua in the complex $\Lambda$ plane.} 
with gauge group $SU(2)$, 
where we identify the quantum numbers $S$ and $T$ correspond to 
the electric and magnetic charges respectively. 
Interestingly, the BPS charge lattice which is 
an abstraction in the four-dimensional context, acquires direct physical 
significance as a lattice of vacua in field space in the two dimensional 
theory.  
   
We now turn to the BPS soliton spectrum for the models with general $N$ and
$\tilde N$. Some partial results for arbitrary $N$ and $\tilde{N}=0$ 
were given in \cite{nick}.
In the general case (with $r>0$), the central charge of 
the classical theory is given by, 
\begin{equation}  
Z=Z_{S}+Z_{T}=i\sum_{i=1}^{N}m_i(S_i+\tau T_{i}) 
+i\sum_{\tilde{i}=1}^{\tilde{N}}\tilde{m}_
{\tilde{i}}
\tilde{S}_{\tilde{i}}
\elabel{cc2}
\end{equation}
In addition to the spectrum of BPS saturated elementary particles described 
above, there are BPS solitons interpolating between each pair of vacua 
${\cal V}_{l}$ and ${\cal V}_{k}$. The (time-independent) soliton solutions are obtained
by a simple embedding of the $CP^1$ soliton (\ref{cov2}) and (\ref{dyon}) by taking
\begin{equation}
w_k^{(l)}(x)=\exp\,\pm|m_l-m_k|x
\elabel{solit}
\end{equation}
where $w^{(l)}_{k}$ is the scalar component of the superfield 
$W^{(l)}_{k}$ introduced above. 
This soliton carries topological charge $T_{i}=\pm\delta_{il} \mp
\delta_{ik}$. Each of these solitons yields an infinite tower of dyons 
which also carry global charges $S_{i}=ST_{i}$ with integer
$S$. However, this is not the complete story since the soliton can
also form bound-states with the various fermions in the theory. The soliton
has an associated fermion $\psi_k^{(l)}$, the superpartner of
$w_k^{(l)}$, which possesses two zero-modes in the background of the soliton 
(\ref{solit}), this is true even in the $CP^1$ case
\cite{nick}. These zero-modes are associated to fermionic creation
operators which create states that fill out the BPS supermultiplet of
the soliton. In the $CP^1$ model this is the complete story; however,
for the more general models there are more fermion fields and the
possibility of more bound-states exists.

In particular, the (time-independent) Dirac equation of the
two-component fermion fields $\psi_j^{(l)}$, $j\neq k$,
and $\tilde \psi_{\tilde j}^{(l)}$ is non-trivial in the soliton
background:
\begin{equation}
\begin{pmatrix} 
|m'|\sin\alpha & i\partial_x+i|m'|\cos\alpha-2i|m|\rho^{-1}w^2\\
i\partial_x-i|m'|\cos\alpha & -|m'|\sin\alpha \end{pmatrix}
\psi=E\psi
\elabel{dirac}\end{equation}
where $\alpha={\rm arg}(m'/m)$. 
In the above, $w=\exp|m|x$ is the soliton solution, with topological
charge $T_i=\delta_{il}-\delta_{ik}$, $m=m_l-m_k$
and $\psi$ is one of the fermion fields 
$\psi_j^{(l)}$,
or $\tilde \psi_{\tilde j}^{(l)}$, with $m'$ equal to $m_l-m_j$ and $
m_l-\tilde m_{\tilde j}$, respectively. The equation admits a normalizable solution,
namely
\begin{equation}
\psi=\begin{pmatrix} 1 \\ 0\end{pmatrix}\exp(|m'|\cos\alpha\,x)
\end{equation}
with energy $E=|m'|\sin\alpha$, as long as
\begin{equation}
0<|m'|\cos\alpha<|m|\, ,\qquad{\rm i.e.}\qquad 0<{\rm Re}\left(\frac{m'}{m}\right)
<1
\end{equation}
In the standard picture of semi-classical quantization, this
normalizable mode is then associated
with a fermion creation and annihilation operator, $\rho$ and
$\rho^\dagger$, with the usual anti-commutation relation
$\{\rho,\rho^\dagger\}=1$. The soliton carries
the two-dimensional Fock space representation of these operators,
i.e. it can exist in the vacuum state $|0\rangle$, the original soliton, and the
bound-state $\rho^\dagger|0\rangle$. Since the fermion mode carries
an energy $|m'|\sin\alpha$, the bound-state has a different mass than
the soliton $M=|i\tau m|$. However, the bound-state is still a BPS
state because the fermion mode contributes to the central charge. The argument goes as
follows: the bound-state has central charge 
$Z=i\tau m+im'$ and therefore to be a BPS state the mass should be
\begin{equation}
M_{\rm b.s.}=|i\tau m+m'|=|i\tau
m|+|m'|\sin\alpha+\cdots
\end{equation}
The first term is the mass of the soliton and the second is the energy
of the fermion mode and the higher term are suppressed in $r^{-1}$. So
to leading order in $r^{-1}$, the bound-state is indeed a BPS state. The corrections
in $r^{-1}$ arise from the back-reaction of the fermion field on the
soliton that we have ignored in our leading-order analysis. At the boundaries of the
region where the fermion mode becomes non-normalizable, the bound-state will
decay into a soliton and a fundamental fermion. 

We can now build up the following picture of the spectrum of
topologically charged BPS states. For a soliton of topological charge
$T_i=\delta_{il}-\delta_{ik}$, there is tower of dyon states with
$S_i=ST_i$. In addition, for each state of the tower there are
bound-state with fermions:

(1) With $\psi_j^{(l)}$ in the region
\begin{equation}
0<{\rm Re}\left(\frac{m_l-m_j}{m_l-m_k}\right)<1
\label{reg1}\end{equation}
The bound-state has $S_i=ST_i+\delta_{il}-\delta_{ij}$ and $\tilde
S_{\tilde i}=0$. At the boundary ${\rm Re}(m_l-m_j/m_l-m_k)=0$ the
bound-state decays to the soliton with $S_i=ST_i$ and the fundamental
fermion with $S_i=\delta_{il}-\delta_{ij}$ and $\tilde
S_{\tilde i}=0$, while at the other
boundary ${\rm Re}(m_l-m_j/m_l-m_k)=1$ the
bound-state decays to the soliton with $S_i=(S+1)T_i$ and the fundamental
fermion with $S_i=\delta_{ik}-\delta_{ij}$ and $\tilde
S_{\tilde i}=0$.
 
(2) With $\tilde\psi_{\tilde j}^{(l)}$ in the region
\begin{equation}
0<{\rm Re}\left(\frac{m_l-\tilde m_{\tilde j}}{m_l-m_k}\right)<1
\label{reg2}\end{equation}
The bound-state has $S_i=ST_i+\delta_{il}$ and $\tilde
S_{\tilde i}=-\delta_{\tilde i\tilde j}$. At the boundary ${\rm
Re}(m_l-\tilde m_{\tilde j}/m_l-m_k)=0$ the
bound state decays to the soliton with $S_i=ST_i$ and the fundamental
fermion with $S_i=\delta_{il}$ and $\tilde S_{\tilde i}=-\delta_{\tilde
i\tilde j}$, while at the other
boundary ${\rm Re}(m_l-\tilde m_{\tilde j}/m_l-m_k)=1$ the
bound state decays to the soliton with $S_i=(S+1)T_i$ and the fundamental
fermion with $S_i=\delta_{ik}$ and $\tilde S_{\tilde i}=-\delta_{\tilde i\tilde j}$.

In overlapping regions (\ref{reg1}) and (\ref{reg2}), for a number of
different fermions $\{j_1,\ldots,j_p,\tilde j_1,\ldots,\tilde j_s\}$ there
will be multiple bound-states carrying global charges
\begin{equation}
S_i=ST_i+(p+s)\delta_{il}-\delta_{ij_1}-\cdots-\delta_{ij_p}\ ,\qquad
\tilde S_{\tilde i}=-\delta_{\tilde i\tilde j_1}-\cdots-\delta_{\tilde
i\tilde j_s}
\elabel{mbs}\end{equation}
 


\section{Theory A: Quantum Effects}
 
In this section we consider how the classical BPS spectrum of Theory A, 
obtained in the previous section, is modified by quantum corrections. 
One particularly important quantum effect, which arises at one loop, 
is the logarithmic running of the of the FI parameter. When all twisted 
masses are set to zero,   
\begin{equation}
r(\mu)=r_{0}-\frac{(N-\tilde{N})}{4\pi}
\log\left(\frac{M_{UV}^{2}}{\mu^{2}}\right)
\elabel{renorm2}
\end{equation}
where $r(\mu)$ is the renormalized FI parameter, defined at the scale 
$\mu$, and $M_{UV}$ is a UV cut-off.  
The bare FI parameter, $r_{0}$, is equal to the renormalized 
FI parameter, $r(\mu)$, 
evaluated at the cut-off scale $\mu=M_{UV}$. 
In the limit $e\rightarrow \infty$, the theory reduces to a supersymmetric 
$\sigma$-model with target space ${\cal M}_{H}$ as described in the 
previous section. The running coupling of the 
$\sigma$-model is 
related to the FI parameter as $g(\mu)=\sqrt{2/r(\mu)}$. 
We will assume that $r_{0}\gg0$ so that the resulting 
$\sigma$-model is weakly-coupled at the cutoff scale. As usual the running 
coupling may be eliminated in favour of an RG invariant scale defined by, 
\begin{equation} 
\Lambda=\mu\exp\left(-\frac{2\pi r(\mu)}{N-\tilde{N}} \right)
\elabel{lam2}
\end{equation}
 
The sign of the $\sigma$-model $\beta$-function  
depends on the sign of $N-\tilde{N}$. If $N>\tilde{N}$ then the theory 
is asymptotically free and, in the absence of twisted masses, 
will run to strong coupling at low energies. This, for example, is 
the behaviour of the supersymmetric $CP^{N-1}$ $\sigma$-model. 
As explained in \cite{nick}, this behaviour is modified when non-zero 
twisted masses are introduced. 
The running of the coupling is driven by quantum fluctuations 
of the light chiral multiplets which appear in the $\sigma$-model 
Lagrangian. All these fields decouple at energy scales below some scale $M$ 
which is roughly the twisted mass of the lightest chiral multiplet. 
The running coupling is therefore frozen below this energy scale. 
Provided we choose twisted masses such that $|m_{i}-m_{j}|\gg\Lambda$ 
for enough pairs $i$ and $j$, 
then the resulting $\sigma$-model will be weakly coupled at all 
energies. On the other 
hand, if $N<\tilde{N}$, then the $\sigma$-model is IR free. Such a theory 
may however suffer from a Landau pole in the UV, and should usually only be 
considered as a low-energy effective theory valid below the mass scale 
$\Lambda$ which may be much lower than $M_{UV}$. Finally, the massless 
theory with $N=\tilde{N}$ has vanishing beta-function and is therefore 
scale invariant. We will henceforth 
restrict our attention to the asymptotically free case $N>\tilde{N}$ 
unless otherwise stated.  
 
Another important effect which arises at one-loop is an anomaly 
which breaks the $U(1)_A$ R-symmetry down to $Z_{2N-2\tilde{N}}$. 
In the absence of twisted masses this means that the bare $\theta$-parameter 
can be set to zero by a $U(1)_{A}$ rotation of the fields. However, 
even if the bare parameter is set to zero the theory will still have an 
effective vacuum angle at low energy. If the twisted masses are non-zero 
then the $U(1)_{A}$ symmetry is already explicitly broken to $Z_{2}$ at 
the classical level. As discussed in \cite{nick}, the one-loop 
effects of the running coupling and the anomaly are equivalent to 
a holomorphic renormalization of the complex coupling $\tau$. In the 
$CP^{1}$ case described in the previous section this amounts to replacing 
$\tau$ by,  
\begin{equation}
\tau_{\rm eff}=ir_{\rm eff}+\theta_{\rm eff}/2\pi=
\frac{i}{\pi }\log\left(\frac{m}{\Lambda}\right)
\elabel{tau1loop}
\end{equation}
When this replacement is combined with the $\theta$-dependent shift in 
global $U(1)$ charge of the dyons described above, it leads to a 
non-trivial monodromy in the dyon spectrum which is a close analog 
of the weak-coupling monodromy of the BPS spectrum of an ${\cal N}=2$ 
theory in four dimensions. 
 
So far we have only discussed the spectrum in the IR limit $e\rightarrow 
\infty$ when the gauge theory we started with reduces to a non-linear 
$\sigma$-model. As long as the twisted masses are large and the 
$\sigma$-model is weakly coupled, this limit is 
convenient for determining the BPS spectrum. However for more general values 
of the twisted masses, the $\sigma$-model becomes strongly coupled and 
a new approach is required. A key property of theories with ${\cal N}=(2,2)$ 
supersymmetry in two dimensions is that the masses of BPS states are 
effectively determined by the F-terms in the superspace Lagrangian. As the 
two dimensional gauge coupling $e$ only appears in the gauge multiplet 
kinetic term, which is a D-term, it follows that the masses of BPS states 
are actually independent of $e$. This suggests a completely different regime 
in which we may attempt to determine the BPS spectrum. When $e$ is much 
less than the other mass scales in the problem we should be able to 
determine the BPS spectrum using ordinary perturbation theory. Note that 
perturbation theory in the two-dimensional gauge coupling is quite distinct 
from the perturbation theory in the $\sigma$-model coupling used above.   
 
Following \cite{witn=2}, we consider the effective Lagrangian along 
the `Coulomb branch' $\phi_{i}=\tilde{\phi}_{\tilde{i}}=0$ with $\sigma$ 
unconstrained. As discussed in the previous section, the are no classical 
SUSY vacua on this branch unless $r=0$. However, even at weak 
coupling, this conclusion can be altered by quantum effects.   
As long as 
$e$ is much less than each of the scales $|\sigma+m_{i}|$, 
$|\sigma +\tilde{m}_{\tilde{i}}|$ and $\Lambda$, we have a theory of a 
light $U(1)$ gauge multiplet weakly coupled to massive chiral multiplets.  
In this case we may integrate out the chiral multiplets and get an effective 
Lagrangian for the gauge degrees of freedom of the form,   
\begin{equation} 
{\cal L}_{\rm eff}=\int  d^{4}\vartheta \,\, {\cal K}_{\rm eff}[\Sigma,
\Sigma^{\dagger}] + \int d^{2}\vartheta\,\, {\cal W}_{\rm eff}[\Sigma] +
\int d^{2}\bar{\vartheta}\,\,\bar{{\cal W}}_{\rm eff}[\Sigma^{\dagger}]
\elabel{leff}
\end{equation}
A one-loop calculation of the effective twisted superpotential yields 
\cite{hh},  
\begin{equation}
{\cal W}_{\rm eff}=\frac{i}{2}\hat{\tau}\Sigma-\frac{1}{4\pi }
\sum_{i=1}^{N}(\Sigma+m_{i})\log\left(\frac{2}{\mu}
(\Sigma+m_{i})\right)+\frac{1}{4\pi }
\sum_{\tilde{i}=1}^{\tilde{N}}(\Sigma+\tilde{m}_{\tilde{i}})
\log\left(-\frac{2}{\mu}
(\Sigma+\tilde{m}_{\tilde{i}})\right)
\elabel{weff2}
\end{equation}
Here, the complexified coupling constant 
$\hat{\tau}$ is equal to $ir(\mu)+\theta/2\pi+n^{*}$ where the integer $n^{*}$ 
is chosen to minimize the potential energy. As explained in \cite{witn=2}, 
this minimization of the potential reflects the fact 
that a non-zero value of the $\theta$ parameter in two-dimensions corresponds 
to a constant background electric field \cite{COL}. The states of the system 
with $n\neq n^{*}$, are unstable to pair creation of 
charged particles which screens the background field leaving the state with 
$N=n^{*}$. In fact there are various arguments which suggest that 
this is the {\em exact} twisted superpotential \cite{DDDS,CV2}.  
In any case, we will always take $e$ sufficiently small so that 
the twisted superpotential is reliable. In general the 
only points where this effective description may break down even at 
small $e$ are the points 
$\sigma=-m_{i}$ and $\sigma=-\tilde{m}_{\tilde{i}}$ 
where we will find that 
extra light degrees of freedom must be included. 
 
The potential energy of the effective theory is, 
\begin{equation}
U=g^{\Sigma\bar{\Sigma}}\left
|\frac{\partial{\cal W}_{\rm eff}}{\partial\Sigma}\right|^{2}
\elabel{pe2}
\end{equation}
where $g^{\Sigma\bar{\Sigma}}=(g_{\Sigma\bar{\Sigma}})^{-1}$ is the 
inverse of the K\"{a}hler metric, 
\begin{equation}
g_{\Sigma\bar{\Sigma}}=-\frac{\partial^{2}{\cal K}_{\rm eff}}
{\partial\Sigma\partial
\Sigma^{\dagger}}
\elabel{eeff} 
\end{equation}
At weak coupling, corrections to the tree level K\"{a}hler potential 
${\cal K}_{tree}=\Sigma^{\dagger}\Sigma/4e^{2}$, are small an we can safely 
assume that the metric does not have poles or zeros in the region of field 
space where our approximations are valid. Thus the supersymmetric vacua 
of the theory are in one to one correspondence with the zeros of 
$\partial {\cal W}/\partial\Sigma$ and are therefore 
determined by the complex equation,  
\begin{equation}
\frac{\prod_{i=1}^{N}(\sigma+m_{i})}
{\prod_{\tilde{i}=1}^{\tilde{N}}(\sigma+\tilde{m}_{\tilde{i}})} 
=\tilde{\Lambda}^{N-\tilde{N}}
\elabel{vacuumeq2}
\end{equation}
where 
$\tilde{\Lambda}=\ft12(-1)^{\tilde{N}/(N-\tilde{N})}\Lambda\exp(-1+i\theta/(N-\tilde{N}))$. 
Provided the numerator and denominator on the LHS of (\ref{vacuumeq2}) do not 
have common zeros, the equation becomes, 
\begin{equation}
\prod_{i=1}^{N}(\sigma+m_{i})-\tilde{\Lambda}^{N-\tilde{N}}
\prod_{\tilde{i}=1}^{\tilde{N}}(\sigma+\tilde{m}_{\tilde{i}}) 
= \prod_{i=1}^{N}(\sigma-e_{i})=0
\elabel{vacuumeq3}
\end{equation}
Thus there are $N$ supersymmetric vacua located at 
$\sigma=e_{i}$ for $i=1,\dots,N$. As above, we have assumed that 
$N>\tilde{N}$. For $|m_{i}-m_{j}|\gg\Lambda$, these vacua coincide with the 
$N$ classical vacua ${\cal V}_{i}$, located at the points $\sigma=-m_{i}$, 
defined in the previous section.     
 
We now consider a BPS saturated soliton obeying the boundary conditions, 
$\sigma \rightarrow e_{k}$ as $x\rightarrow -\infty$ and 
$\sigma \rightarrow e_{l}$ as $x\rightarrow +\infty$. 
As the effective Lagrangian is of the standard Landau-Ginzburg form 
discussed in the previous section we may apply the general BPS mass 
formula obtained there. 
Thus, the soliton mass is given by $M_{kl}=|Z_{kl}|$ where 
$Z_{kl}=2\Delta{\cal W}_{\rm eff}=2{\cal W}_{\rm eff}(e_{l})
-2{\cal W}_{\rm eff}(e_{k})$. A short calculation reveals that, 
\begin{eqnarray} 
Z_{kl}& = & \frac{1}{2\pi}\left[(N-\tilde{N})(e_{l}-e_{k})-
\sum_{i=1}^{N}m_{i}\log\left(\frac{e_{l}+m_{i}}
{e_{k}+m_{i}}\right)+\sum_{\tilde{i}=1}^{\tilde{N}}\tilde{m}_{\tilde{i}}
\log\left(\frac{e_{l}+\tilde{m}_{\tilde{i}}}
{e_{k}+\tilde{m}_{\tilde{i}}}\right)\right]
\elabel{zkl}
\end{eqnarray}
For non-zero twisted masses, the 
branch-cuts in the logarithms appearing in (\ref{zkl}) 
lead to an ambiguity in the BPS 
spectrum. In particular the ambiguity in the central charge is equal to 
$i\sum_{i=1}^{N}m_{i}n_{i}+i\sum_{\tilde{i}=1}^{\tilde{N}}
\tilde{m}_{\tilde{i}}\tilde{n}_{\tilde{i}}$ where the choice of integers 
$n_{i}$ and $\tilde{n}_{\tilde{i}}$ corresponds to 
a choice of branch for each of the $N+\tilde{N}$ logarithms in (\ref{zkl}). 
As explained by Hanany and Hori in \cite{hh}, this ambiguity signals the 
fact that solitons can 
carry integer values of the global $U(1)$ charges $S_{i}$, 
$\tilde{S}_{\tilde{i}}$ in addition to 
their topological charges. This is related to the 
existence of BPS dyons at weak-coupling discussed in the 
previous sections. Including this effect, the final formula for the 
masses of all BPS states in the theory is $M=|Z|$ with,  
charge is 
\begin{equation}  
Z=i\sum_{i=1}^{N}(m_iS_i+ m_{Di}T_{i}) 
+i\sum_{\tilde{i}=1}^{\tilde{N}}\tilde{m}_{\tilde{i}}
\tilde{S}_{\tilde{i}}
\elabel{cc4}
\end{equation} 
which  differs from the classical 
formula (\ref{cc2}) of the previous section by the replacement of 
$m_{Di}^{cl}=\tau m_{i}$ by 
\begin{equation}
m_{Di}=-2i{\cal W}_{\rm eff}(e_{i})=\frac{1}{2\pi i}\left[(N-\tilde{N})e_{i}- 
\sum_{j=1}^{N}m_{j}\log(e_{i}+m_{j})+
\sum_{\tilde{j}=1}^{\tilde{N}}\tilde{m}_{\tilde{j}}
\log(e_{i}+\tilde{m}_{\tilde{j}}) \right]
\elabel{md} 
\end{equation} 
For $|m_{i}-m_{j}|\gg\Lambda$, this formula 
can be directly 
compared with the results of the semiclassical analysis given 
in the previous sections. In this limit, the replacement 
$\tau\rightarrow \tau_{\rm eff}=\partial {\cal W}/\partial \sigma$ 
reproduces the one-loop holomorphic renormalization of $\tau$, 
given above by (\ref{tau1loop}) in the $CP^{1}$ case, 
while in the general case the semi-classical limit is
\begin{equation}
m_{Di}\rightarrow\frac{1}{2\pi i}\left[-(N-\tilde{N})m_{i}- 
\sum_{j=1}^{N}(m_{j}-m_i)\log\left(\frac{m_{j}-m_i}{\tilde\Lambda}\right)+
\sum_{\tilde{j}=1}^{\tilde{N}}(\tilde{m}_{\tilde{j}}-m_i)
\log\left(\frac{\tilde{m}_{\tilde{j}}-m_i}{\tilde\Lambda}\right)\right]
\elabel{scmass} 
\end{equation} 
 
For particular values of the twisted masses there are various singular 
points which can be compared with the corresponding singular points 
in the classical theory. For example, in the previous section we discovered 
that if two twisted masses $m_{i}$ and $m_{j}$ coincide 
in the classical theory, then the chiral multiplets $\phi_{i}$ and $\phi_{j}$ 
are both massless at $\sigma=-m_{1}=-m_{2}$ giving 
a Higgs branch which is a copy of $CP^{1}$. 
As the resulting low-energy $\sigma$-model is strongly 
coupled in the IR we should expect that quantum effects modify this 
singularity. As explained in \cite{nick}, the classical 
singular point $m=m_{i}-m_{j}=0$, is split into a pair of singular points at 
which the vacua $e_{i}$ and $e_{j}$ coincide. At each of these 
singular points a single chiral multiplet, corresponding to the 
soliton which interpolates between the vacua 
${\cal V}_{i}$ and ${\cal V}_{j}$, becomes massless. A single massless 
chiral multiplet does not yield a continuous Higgs branch and, 
in particular, does not result in a massless Goldstone boson. Thus 
there is no conflict with the usual restrictions on massless particles in 
two dimensions \cite{Col2}. 
 
In the present case we can also consider what 
happens if chiral multiplets of opposite charge become massless. In the 
classical theory this happens if $m_{i}=\tilde{m}_{\tilde{j}}$ for some 
$i$ and $\tilde{j}$. As in the previous case this gives a Higgs branch 
of complex dimension one at $\sigma=-m_{i}=-\tilde{m}_{\tilde{j}}$. 
However the present Higgs branch has zero first Chern class and the 
corresponding low-energy $\sigma$-model is scale-invariant. In this 
case we might reasonably expect the Higgs branch to remain in the 
full quantum theory. The massless theory on the Higgs branch could either 
be a free theory or, perhaps, a non-trivial CFT\footnote{Neither of these 
possibilities is in conflict with the absence of Goldstone bosons in 
two-dimensions.}. In equation (\ref{vacuumeq2}), which determines the 
exact location of the Coulomb branch vacua, something interesting happens 
whenever $m_{i}=\tilde{m}_{\tilde{j}}$ for some $i$ and $\tilde{j}$: 
the numerator and denominator have a common zero and a cancellation occurs. 
The degree of the polynomial (\ref{vacuumeq3}) is then lowered by one and 
it appears that one of the supersymmetric vacua has disappeared. If this were 
true it would, certainly contradict standard properties of the Witten index. 
The resolution is simply that, as in the classical theory, a non-compact 
Higgs branch appears at $\sigma=-m_{i}$. The effective FI parameter 
on this branch is determined by the twisted superpotential,  
\begin{equation}
\tau_{\rm eff}=\left. \frac{\partial{\cal W}}{\partial \sigma} \right|_
{\sigma=-m_{i}} 
\elabel{FIeff}
\end{equation}
It is also useful to note that, as the two massless chiral multiplets 
$\Phi_{i}$ and $\tilde{\Phi}_{j}$ have opposite charges, we can introduce 
a complex mass $\hat{m}_{i\tilde{j}}$ by including 
a superpotential of the form (\ref{complexm}). This lifts the Higgs 
branch completely and, in this case, we are left with only $N-1$ Coulomb 
branch vacua. This is particularly clear in the IIA brane picture described 
in Section 5. 
A contradiction with the Witten index is avoided because the 
transition from theory with $N$ vacua to one with $N-1$, always goes via 
the singular point in  parameter space where 
$m_{i}=\tilde{m}_{\tilde{j}}$, $\hat{m}_{i\tilde{j}}=0$. At this point 
we have flat directions on the Higgs branch and the Witten index is not 
defined. 
 
The masses of topologically non-trivial BPS states are patently not
single valued as one tracks their mass, as a function of the
parameters $m_i$ and $\tilde m_{\tilde i}$, around one of the
singularities described above: there is non-trivial monodromy. 
To make this concrete we can choose a
particular branch of the multi-valued functions (\ref{scmass}) and
then interpret the monodromy as transformations on the charges as one
crosses a cut. Around the singularity $m_j=m_l$ the transformation is
\begin{equation}
T_i\rightarrow T_i,\qquad S_i\rightarrow
S_i+\delta_{il}-\delta_{ij},\qquad
\tilde S_{\tilde i}\rightarrow\tilde S_{\tilde i}
\elabel{monod1}\end{equation}
while around the second kind of singularity $\tilde m_{\tilde j}=m_l$
\begin{equation}
T_i\rightarrow T_i,\qquad S_i\rightarrow S_i+\delta_{il},\qquad
\tilde S_{\tilde i}\rightarrow\tilde S_{\tilde
i}-\delta_{\tilde i\tilde j}
\elabel{monod2}\end{equation}
It is easy to see that if $j\neq k$ in (\ref{monod1}) and for any
$\tilde j$ in (\ref{monod2}), repeated application of these transformations does not 
preserve the semi-classical spectrum. For instance, for the states
with $T_i=\delta_{il}-\delta_{ik}$ and $S_i=ST_i$, for integer $S$,
the monodromy transformation around the singularity $m_j=m_l$, $j\neq k,l$, induces
$S_i\rightarrow ST_i+\delta_{il}-\delta_{ij}$, which is a state in the
spectrum in the region (\ref{reg1}). However, a further such transformation
gives $S_i\rightarrow ST_i+2\delta_{il}-2\delta_{ij}$ which is not 
commensurate with the semi-classical spectrum. The resolution of this
paradox, is that the contour around the singularity $m_j=m_l$ passes
through a curve of marginal stability on which the state decays into
two other topologically charged states with charges
$T_i^{(1)}=\delta_{il}-\delta_{ij}$, $S^{(1)}_i=(S+2)T_i^{(1)}$ and 
$T_i^{(2)}=\delta_{ij}-\delta_{ik}$, $S^{(2)}_i=ST_i^{(2)}$. Notice
that this kind of decay is in addition to the soliton decay to soliton
plus fundamental fermion that we saw in the last section. If we
think of the masses $m_i$ and $\tilde m_{\tilde i}$ as a set of points
in the complex plane, then the curve of marginal stability is
identified with configurations of points where $m_j$ lies on the
straight line joining $m_k$ and $m_l$. On the curve, the simple
semi-classical analysis of section 2 must be re-evaluated because the
moduli space of the charge $T_i=\delta_{il}-\delta_{ik}$ soliton is
larger than that of the embedded $CP^1$ soliton. The enlarged moduli
space of solutions, can be interpreted as two superposed solitons with charges
$T_i^{(1)}$ and $T^{(2)}_i$, above, and the original soliton decays at
threshold into these two constituents. This mechanism for decay is precisely
the same as that seen for dyons at weak coupling in supersymmetric
gauge theories \cite{tim,tim2}. 
 
So far we have determined the exact mass formula for BPS states with 
arbitrary global and  topological charges. However, except in the 
weak-coupling limit of the low-energy $\sigma$-model, we do not know which 
states are present in the theory. In general, as we vary the 
parameters appearing in the twisted superpotential, we may cross curves 
of marginal stability on which BPS states can decay. Even in the simplest 
case of target space $CP^{1}$, the existence of such curves was demonstrated 
in \cite{nick}. Fortunately there is a special point in parameter space 
where we may determine the exact BPS spectrum.  We first consider the case 
$\tilde{N}=0$. If we set all twisted masses to zero, we have 
the standard supersymmetric $CP^{N-1}$ $\sigma$-model. This theory 
is integrable and its exact spectrum and S-matrix are known. Also, at this 
point only, the model has an unbroken $SU(N)$ global symmetry 
and the BPS states are organised in multiplets of this symmetry. 
To make contact 
with the analysis of the twisted superpotential we note that, for 
$\tilde{N}=0$ and $m_{i}=0$ the complex equation (\ref{vacuumeq2}) reduces 
to $\sigma^{N}=\tilde{\Lambda}^{N}$. The vacua $e_{k}$, 
$k=1,2,\ldots,N$ are then located at the vertices of a regular $N$-gon 
in the complex $\sigma$-plane, 
\begin{equation}
e_{k}=\tilde{\Lambda}\exp\left(\frac{2\pi ik}{N}\right)
\elabel{sigma1}
\end{equation}
As the twisted masses vanish, the theory has a $Z_{2N}$ symmetry which is 
spontaneously broken to $Z_{2}$ by the VEV of $\sigma$. The $Z_{N}$ symmetry 
of the $N$-gon comes from the quotient $Z_{2N}/Z_{2}$.  
 
According to (\ref{zkl}), the mass of a soliton interpolating between the 
vacua with $\sigma=e_{l}$ and $\sigma=e_{k}$ is 
$M_{k,l}=|2\Delta{\cal W}_{\rm eff}|$. Because of the $Z_{N}$ symmetry 
this only depends on the difference $p=l-k$. The resulting mass is 
\begin{equation}
M_{k,l}=\frac{N}{\pi}\left|\exp\left(\frac{2\pi ip}{N}\right)-1\right|
\tilde{\Lambda} 
\elabel{m=0spectrum}
\end{equation}
It is also known \cite{CV2} that for $p=1,2\ldots N$ the degeneracy 
of BPS states is, 
\begin{equation}
D_{p}=\left(\begin{array}{c} N \\ 
                             p \end{array}\right) 
\elabel{degen}
\end{equation}   
The lightest soliton states with $p=1$ and degeneracy $N$ are interpreted 
as the elementary quanta of the the chiral fields $\Phi_{i}$ \cite{witcpn}. 
In fact they carry charge $+1$ under the unbroken $U(1)$ gauge symmetry.  
These states transform in the fundamental 
representation of the flavour symmetry group $SU(N)$. 
The states with $p>1$ correspond 
to stable bound states of $p$ different flavours of 
elementary quanta and therefore transform in the $p$'th antisymmetric 
tensor representation of $SU(N)$ which agrees with the degeneracy $D_{p}$ in 
(\ref{degen}). These are the only BPS states of the model. 
The complete BPS spectrum of the supersymmetric $CP^{N-1}$ 
$\sigma$-model obtained in this way, including both the 
fundamental solitons and their boundstates, 
is consistent with various exact results which can be obtained by invoking 
the integrability of the model. 
 
In the vicinity of the strong coupling point, where
all the masses $m_i$ are small, the degeneracy of the states is
broken and the states with a topological charge
$T_i=\delta_{il}-\delta_{ik}$ are associated to $D_p$ different
contours in the $\sigma$-plane which join the points $e_l$ and $e_k$.
As explained in \cite{hh}, the different contours can be
constructed from some base contour joining $e_l$ and $e_k$, 
by taking the contours which differ from this by 
encircling, in a positive sense, the $p$ 
distinct points $m_{j_a}$, $a=1,\ldots,p$, chosen from the set
$\{m_i\}$. This gives rise to precisely $D_p$ different states with  
global charge
\begin{equation}
S_i=N_i-\delta_{ij_1}-\cdots-\delta_{ij_p}
\elabel{scgc}\end{equation}
where $N_i$ is some fixed set of integers which depends upon the choice of
branch for the multi-valued function $m_{Di}$. This preferred set of
contours can be shown to arise very naturally from the Type IIA string
description of the model \cite{hh}.
 
We can now compare the strong coupling and weak coupling spectra. It
is clear that, as in the $CP^1$ case, there are an a infinite number of
states at weak coupling while the set of states at strong coupling is
finite. This implies that semi-classical regime must be separated from
the strong coupling regime by curves on which almost
all of the weak coupling states decay. This is familiar from the behaviour of dyon
states in $SU(2)$ ${\cal N}=2$ SQCD \cite{bf}. Moreover, as in gauge
theory case, the strong coupling states
are a subset of the weak coupling states since states with
global charge (\ref{scgc}) are
realized in some region of the moduli space at weak coupling, as
bound-states of the soliton with fermions (\ref{mbs}).
 
Finally, we will consider the theory without twisted masses in the 
more general case of arbitrary $N$ and $\tilde{N}$. In this case there 
is a cancellation of a factor of $\sigma^{\tilde{N}}$ between the numerator 
and denominator of (\ref{vacuumeq2}). As discussed above this 
indicates the presence of a non-compact Higgs branch at $\sigma=0$. 
As $\tilde{N}$ 
chiral multiplets of each charge become massless at this point the 
Higgs branch has complex dimension $2\tilde{N}-1$. After the cancellation the 
equation which determines 
the supersymmetric vacua reduces to $\sigma^{N-\tilde{N}}
=\tilde{\Lambda}^{N-\tilde{N}}$. This 
is the same equation which arises in the case of the 
supersymmetric $CP^{N-\tilde{N}-1}$ $\sigma$-model with 
zero twisted masses.     
This is not a coincidence; we can lift the Higgs branch 
by introducing suitable complex mass terms. As complex masses can be thought 
of as the scalar components of background chiral superfields they cannot 
modify the twisted superpotential which determines the soliton masses. 
By taking the complex masses large we can decouple the $\tilde{N}$ chiral  
multiplets whose VEVs parametrize the Higgs branch and the resulting theory 
theory has $N-\tilde{N}$ chiral multiplets of charge $+1$ under 
$U(1)_{G}$ and none of charge $-1$. The theory then reduces to the  
supersymmetric $CP^{N-\tilde{N}}$ $\sigma$-model at low-energy. As for
the $\tilde N=0$ models discussed above, the strong coupling
spectrum contains a subset of the states that appear at weak coupling.
 
We will now briefly summarize our results for Theory A. The exact 
central charge of the theory is defined by (\ref{cc4}) and (\ref{md}). 
This formula is valid for all $e$ and it applies 
equally in the $\sigma$-model limit $e\rightarrow \infty$. 
If enough of the twisted mass differences $|m_{i}-m_{j}|$ are much greater 
than $|\Lambda|$, the low-energy $\sigma$-model is weakly coupled 
and the BPS spectrum is essentially that of the classical theory described 
in the previous section. In particular the BPS spectrum in the region consists 
of elementary particles, 
kinks which interpolate between each pair of vacua, and an 
infinite tower of dyons associated with each kink. There are also 
soliton-fermion boundstates.  
On the other hand we 
have also determined the BPS spectrum for the theory 
with zero twisted masses. In this case we find, $2\tilde{N}-1$ massless chiral 
multiplets on the Higgs branch while on the Coulomb branch there are 
$N-\tilde{N}$ supersymmetric vacua and a spectrum of massive solitons 
identical to that of the supersymmetric $CP^{N-\tilde{N}-1}$ model. 
The model with zero twisted masses has 
a global $SU(N-\tilde{N})\times U(\tilde{N})$ symmetry and non-anomalous  
$Z_{2N-2\tilde{N}}$ R-symmetry which is spontaneously broken to $Z_{2}$. 
Introducing complex masses lifts the Higgs branch but does not change the 
Coulomb branch vacua or the BPS soliton mass. On the other hand,        
introducing non-zero twisted masses 
breaks the global $SU(N-\tilde{N})$ symmetry. If the masses are small 
the only effect is to introduce small mass splittings in the 
multiplets described above. However, as the masses are increased   
the theory can cross curves of marginal stability on which the BPS spectrum 
can change. In fact, the presence of such curves is certainly 
required to explain the difference between the strong coupling spectrum obtained here 
which has a finite number of BPS states and 
the weak coupling spectrum which has an infinite number. 
\section{Theory B}
 
We turn now to the spectra of the four-dimensional theories described 
in the introduction. Hence we will consider ${\cal N}=2$ SQCD with 
$N$ colours and $N_{f}>N$ fundamental hypermultiplets with 
masses $m_{\lambda}$ for $\lambda=1,\ldots,N_f$. 
This theory has received a great deal of attention in recent years and we 
need only be brief. The bosonic field content of Theory B 
consists of an $SU(N)$ gauge field and a single 
complex scalar, $\phi$, transforming in the adjoint representation 
of the gauge group. A further $N_f$ complex scalars, denoted 
$Q_a^\lambda$, with $a=1,\ldots,N$, transform 
in the fundamental representation of the gauge group and 
$N_f$ complex scalars, $\tilde{Q}^a_\lambda$, in the anti-fundamental 
representation. 
 
The moduli space of supersymmetric vacua is found by 
solving the classical D- and F-term equations. 
A detailed description of the 
various branches which occur was given in \cite{aps} and we will now 
summarize the relevant results. The classical 
theory has a Coulomb branch on which 
the adjoint scalar VEV lies in the Cartan subalgebra of the gauge group: 
$\phi={\rm diag }(\phi_1,\ldots,\phi_{N})$ with $\sum_{a=1}^{N}\phi_a=0$. 
The complex variables 
$\phi_{a}$, with $a=1,\ldots,N$, parametrize gauge-inequivalent vacua, 
modulo the Weyl group of $SU(N)$. The theory also has various Higgs 
branches which are characterised as `baryonic' or 
`non-baryonic', depending on whether or not there is a 
non-vanishing VEV of the gauge invariant operators,
 \be
B^{\lambda_1...\lambda_{N}}&=&Q^{\lambda_1}_{a_1}\ldots
Q^{\lambda_{N}}_{a_{N}}
\epsilon^{a_1...a_{N}} \nn \\
\tilde{B}_{\lambda_1...\lambda_{N}}&=&\tilde{Q}^{a_1}_{\lambda_1}\ldots
\tilde{Q}^{a_{N}}_{\lambda_{N}}\epsilon_{a_1...a_{N}}\ . 
\nn\ee
Non-baryonic branches have $B=\tilde{B}=0$ and are parametrized instead by 
the gauge-invariant `mesons' 
$M_{\rho}^{\lambda}=Q_{\rho}^{a}\tilde{Q}^{\lambda}_{a}$. 
The classical Coulomb branch is corrected by quantum effects while the 
Higgs branches, being protected by a non-renormalization theorem, are not. 
The Higgs branches intersect the Coulomb branch at singular points where 
extra degrees of freedom become massless. Although the Higgs branches 
themselves are not corrected, their intersection with the Coulomb branch 
may be different in the classical and quantum theories. 
      
In the present context we are interested in the theory at the 
root of its first baryonic branch. In the classical theory with 
vanishing masses, $m_{\lambda}=0$, this branch has quaternionic dimension 
$N_{f}N-N^{2}+1$ and emanates from the origin of the 
Coulomb branch. When non-zero masses are introduced we must solve the 
modified F-term equations, 
\be
Q_a^\lambda(m_\lambda+\phi_a)=\tilde{Q}^a_\lambda(m_\lambda+\phi_a)=0 \hspace{0.8in}
{\rm No\ summation\  over\  } \lambda\ {\rm or }\ a.
\elabel{beebop}\ee
In order for the scalars to satisfy (\ref{beebop}) while still 
having non-vanishing expectation value for $B$ and 
$\tilde{B}$, the adjoint 
scalar VEVs must take values $\phi_a=-m_{\lambda(a)}$, 
where $\{m_{\lambda(a)}\}$ is a set of $N$ masses chosen from the 
$m_{\lambda}$, $\lambda=1,\ldots,N_{f}$. Up to 
Weyl group transformations, there are $N_{f}$ choose $N$ 
possible sets and each choice gives a baryonic root from which a 
Higgs branch of complex dimension one may emanate.  
To choose a particular root, we will label the masses as 
$m_{\lambda}=m_{i}$ for $\lambda=i=1,\ldots, N$ and $m_{\lambda}=
\tilde{m}_{\tilde{i}}$ with $\lambda=\tilde{i}+N$ for 
$\tilde{i}=1,\ldots, \tilde{N}=N_{f}-N$ and then set 
$\phi_{a}=-m_{i}\delta_{ia}$. This solution is only consistent with the 
tracelessness 
of the adjoint representation of $SU(N)$ if we have $\sum_{i=1}^{N}m_i=0$. 
As discussed in the introduction 
the masses $m_{i}$ and $\tilde{m}_{\tilde{i}}$ 
will be identified with the twisted masses of Theory A which 
carry the same indices. 
 
At a generic point on its Coulomb branch, the 
four-dimensional theory has classical central charge, 
\begin{equation}
Z=\sum_{a=1}^{N}\, \phi_{a}(q_{a}+ 
\tau h_{a}) + 
\sum_{\lambda=1}^{N_{f}}\, m_{\lambda}s_{\lambda} 
\elabel{4dcc1}
\end{equation}
where $\tau=4\pi i/g^{2}+\theta/2\pi$ is the complexified gauge coupling 
and  $q_{a}$, $h_{a}$ and $s_{\lambda}$ are integer valued electric, 
magnetic and global charges respectively. Specializing this expression to 
the baryonic root defined above we may rewrite the central charge as, 
\begin{equation}
Z=\sum_{i=1}^{N}\, m_{i}(S_{i}+ \tau T_{i}) + 
\sum_{\tilde{i}=1}^{\tilde{N}}\, \tilde{m}_{\tilde{i}}\tilde{S}_{\tilde{i}} 
\elabel{4dcc2}
\end{equation}
where we have redefined charges as 
$S_{i}=-s_{a}-q_{a}$ and $T_{i}=-h_{a}$ for 
$i=a=1,\ldots, N$ and $\tilde{S}_{\tilde{i}}=-s_{\tilde{i}+N}$ for 
$\tilde{i}=1,\ldots,\tilde{N}$. Note that the electric charges become 
parallel to a subset of the global charges at the baryonic root and we 
have absorbed them with an integer shift in the definition of the global 
generators. The resulting expression for the classical central charge 
exactly matches its two-dimensional counterpart (\ref{cc2})
 
Let us now consider which values of the charges $S_{i}$, $\tilde{S_{i}}$ and 
$T_{i}$ occur in the four dimensional theory. We will consider 
${\cal N}=2$ vector multiplets and hypermultiplets in turn.    
At a generic point on the Coulomb branch the theory includes $N-1$ massless 
abelian vector multiplets which correspond to the unbroken 
$U(1)^{N-1}$ gauge symmetry. The remaining generators of the gauge group 
give $N(N-1)$ massive vector multiplets which carry 
non-zero electric charges. At 
the baryonic root the charges which arise correspond to $S_{i}=
\delta_{ik}-\delta_{ij}$ for each $k\neq j$ Thus we have $N(N-1)$ vector 
multiplets with distinct masses $|m_{k}-m_{j}|$. 
The spectrum at the baryonic root also contains 
$N$ massless quark hypermultiplets whose scalar components are 
$Q^{\lambda=i}_{a=i}$ and $\tilde{Q}^{\lambda=i}_{a=i}$ for $i=1,\ldots 
N$. These states have non-zero electric charges $q_{a=i}$ and non-zero global 
charges $s_{\lambda=i}$ which cancel to give $S_{i}=0$. 
These massless quarks each acquire non-zero VEVs on 
the baryonic branch itself. 
 
Finally, and most importantly in the present 
context, Theory B also has a diverse spectrum of massive hypermultiplets 
which includes quarks, monopoles and dyons. These massive quark states 
correspond to the remaining components of the hypermultiplets  
which carry electric charges $q_{a=i}$ and global 
charges $s_{\lambda=i}$ and yield non-zero values 
of $S_{i}$ and $\tilde{S}_{\tilde{i}}$. The resulting spectrum includes 
$N(N-1)$ states with charges $S_{i}=\delta_{ik}-\delta_{ij}$ 
and $\tilde{S}_{\tilde{i}}=0$ which have masses 
$|m_{k}-m_{j}|$ as well as $N(N_{f}-N)$ states with 
charges $S_i=\delta_{ik}$ and 
$\tilde{S}_{\tilde{i}}=-\delta_{\tilde{i}\tilde{j}}$ which have masses 
$|m_{k}-\tilde{m}_{\tilde j}|$. The total number of massive quark 
hypermultiplets is 
therefore $N(N_{f}-1)=N(N+\tilde{N}-1)$. This agrees with the classical 
spectrum of elementary particles of Theory A. 
 
 All further BPS states have non-zero magnetic charge. While 
the existence of classical solutions in various topological sectors 
is well known, the existence of quantum bound states is more 
problematic. For systems with a single Higgs field or, 
equivalently, for real adjoint Higgs VEVs, $\phi_a$, this 
requires detailed knowledge of the monopole moduli spaces for 
higher rank gauge groups (see for example \cite{cedholm}). 
However, as first discussed in \cite{tim}, the situation 
simplifies greatly for generic complex VEVs and 
semi-classical quantization reduces to the corresponding 
problem with all fields restricted to $SU(2)$ subgroups 
of the gauge group. The spectrum arising from each 
subgroup can then be determined using a variety 
of methods \cite{gh,ssz,bf}. The resulting monopole 
spectrum contains $N(N-1)$ different states with magnetic 
charges $T_i=\delta_{il}-\delta_{ik}$ for each $l\neq k$ 
and with masses $4\pi |m_l-m_k| /g^2$, in agreement with 
the classical soliton spectrum of Theory A. In addition to these 
purely magnetic states, the spectrum also contains an 
infinite tower of dyons associated to each of the $SU(2)$ 
subgroups and with charges $S_i=ST_i$ for integer values of 
$S$. Once again, this coincides with the excited soliton 
spectrum of Theory A. 

However, just as for Theory A, this is not the
whole story because dyons can form bound-states with quarks. To
our knowledge, the spectrum of such states has not been constructed in
the literature and so to this we devote Appendix B. The results
are as follows. We work at a generic point in the moduli space, so
that the moduli space of a monopole is that of a monopole in an $SU(2)$
gauge theory. At the baryonic root, we find the following spectrum
of bound-states of quarks with 
the dyon having charges $T_i=\delta_{il}-\delta_{ik}$ and $S_i=ST_i$:

(1) With the quark with charges $S_i=\delta_{il}-\delta_{ij}$ and $\tilde
    S_{\tilde i}=0$ in the region
\begin{equation}
0<{\rm Re}\left(\frac{m_l-m_j}{m_l-m_k}\right)<1
\label{reg3}\end{equation}

(2) With the quark with charges $S_i=\delta_{il}$ and $\tilde S_{\tilde i}=-\delta_{\tilde i\tilde
    j}$ in the region 
\begin{equation}
0<{\rm Re}\left(\frac{m_l-\tilde m_{\tilde j}}{m_l-m_k}\right)<1
\label{reg4}\end{equation}

Notice that the bound-states have exactly the same quantum numbers as
the soliton fermion bound-states in Theory A and moreover regions
(\ref{reg3}) and (\ref{reg4}) are precisely the same as regions 
(\ref{reg1}) and (\ref{reg2}), respectively. 
At the boundary the bound-state will decay to a dyon and a quark in
exactly the same way as in Theory A. We conclude from this that the
weak coupling spectrum of topologically charged BPS states is identical in
Theory A and B.

While we have shown classical agreement between the masses of the 
BPS massive hypermultiplet spectrum of Theory B and the 
the BPS spectrum of Theory A, the spectrum and 
low-energy interactions of the   
four-dimensional ${\cal N}=2$ supersymmetric gauge theory are corrected 
by quantum effects. In the asymptotically free case $N_{f}<2N$ the 
coupling constant runs at one loop and the complex classical parameter 
$\tau$ is replaced by the RG invariant scale $\Lambda$. The theory 
also has an anomaly in the $U(1)$ part of the R-symmetry group which is broken 
to $Z_{4N-2N_{f}}$. This residual discrete symmetry is explicitly 
broken down to $Z_{2}$ for non-zero hypermultiplet masses.   
At the baryonic root, Theory B is described by the 
degenerate elliptic curve, 
\be
F(t,v)=\left(t\Lambda^{N-N_f}\prod_{\tilde{i}=1}^{N_f-N}
(v-\tilde{m}_{\tilde{i}})
-\prod_{i=1}^{N}(v-m_i)\right)\left(t-\Lambda^{N}\right)
\elabel{Ftv2}\ee
This is equivalent to the curve describing the baryonic root 
given in \cite{aps}. All parameters and variables in the curve 
have even charge under 
$Z_{4N-2N_{f}}$ and thus, as in \cite{aps}, 
the effective symmetry of the curve is the $Z_{2}$ 
quotient $Z_{2N-N_{f}}=Z_{N-\tilde{N}}$.  
This form of the curve occurs naturally in the 
M-theory construction of \cite{witm5} which we review in Section 5. 
The curve is branched over the $N$ points $e_{i}$, with $i=1,2,\ldots,N$, 
defined by, 
\begin{equation}
\prod_{i=1}^{N}(v+m_{i})-\Lambda^{N-\tilde{N}}
\prod_{\tilde{i}=1}^{\tilde{N}}(v+\tilde{m}_{\tilde{i}}) 
= \prod_{i=1}^{N}(v-e_{i})=0
\elabel{ei4d}
\end{equation}   
Note that this is the same as equation (\ref{vacuumeq3}) 
which determines the critical points of the twisted superpotential of 
Theory A. 
 
As in Theory A, the case of vanishing masses 
$m_{i}=\tilde{m}_{\tilde{i}}=0$ will be 
of particular interest. In this case, 
the low-energy theory at the baryonic root has an unbroken $SU(\tilde{N})
\times U(1)^{N-\tilde{N}}$ 
gauge symmetry \cite{aps} with $N_{f}$ massless flavours in the fundamental 
representation of $SU(\tilde{N})$.  
Thus the spectrum of the theory includes the corresponding 
massless quarks and gluons.  
This is consistent because the  
condition $N_{f}=N+\tilde{N}<2N$ which guarantees that the microscopic
theory 
is asymptotically free, implies that $N_{f}>2\tilde{N}$, which ensures that 
the low-energy theory is IR free. The massless spectrum at this point 
also includes $N-\tilde{N}$ hypermultiplets which are neutral under 
the non-abelian factor of the unbroken gauge group. The theory with zero 
masses also has an unbroken 
$SU(N-\tilde{N})$ global symmetry and 
a $Z_{2N-2\tilde{N}}$ R-symmetry which is unbroken at 
the baryonic root.      
 
In the quantum theory the central charge is given by the integral 
of the Seiberg-Witten 
differential $\lambda_{SW}=vd(\log t)$ 
over certain one cycles of the curve. The 
resulting modification of the classical formula (\ref{4dcc2}) is 
\begin{equation}
Z=\sum_{i=1}^{N}\, (m_{i}S_{i}+ m_{Di}T_{i}) + 
\sum_{\tilde{i}=1}^{\tilde{N}}\, \tilde{m}_{\tilde{i}}\tilde{S}_{\tilde{i}} 
\elabel{4dcc3}
\end{equation}
with 
\begin{equation}
m_{Dl}-m_{Dk}=\frac{1}{2\pi i}\int_{e_{k}}^{e_{l}}d\lambda_{\rm SW}=
\frac{1}{2\pi i}\int_{e_{k}}^{e_{l}} v \frac{dt}{t}
\elabel{swdiff}
\end{equation}
which gives 
\begin{eqnarray}
m_{Dl}-m_{Dk} & = & \frac{1}{2\pi}\left[
\sum_{i=1}^{N}\,\int^{e_{l}}_{e_{k}} \frac{v\,dv}{v+m_{i}} 
- \sum_{\tilde{i}=1}^{\tilde{N}}\,\int^{e_{l}}_{e_{k}} \frac{v\,dv}
{v+\tilde{m}_{\tilde{i}}} \right]
\nonumber \\
& = & \frac{1}{2\pi}\left[  (N-\tilde{N})(e_{l}-e_{k})-
\sum_{i=1}^{N}m_{i}\log\left(\frac{e_{l}+m_{i}}
{e_{k}+m_{i}}\right)+\sum_{\tilde{i}=1}^{\tilde{N}}\tilde{m}_{\tilde{i}}
\log\left(\frac{e_{l}+\tilde{m}_{\tilde{i}}}
{e_{k}+\tilde{m}_{\tilde{i}}}\right) \right] \nonumber \\
\elabel{fr}
\end{eqnarray}
Thus we have shown that the central charge 
of Theory B agrees with the exact central charge (\ref{cc4},\ref{md}) of 
Theory A in the previous section. 
 
Although we have complete agreement between the central charges of 
Theory A and Theory B, it still remains to compare which values of the 
quantum numbers $S_{i}$, $\tilde{S}_{\tilde{i}}$ and $T_{i}$ actually appear 
in the BPS spectra of the two theories. As long as we choose the masses 
$m_{i}$ so that a sufficient number of gauge bosons have masses much larger 
than the dynamical scale $\Lambda$, Theory B is weakly coupled and its 
spectrum is essentially the classical spectrum described above. Hence, 
in this regime, massive hypermultiplets of Theory B are in 
one-to-one correspondence with BPS states of Theory A. 
However, there is more: the curves of marginal stability at weak
coupling in Theory A, where
topologically charged BPS solitons decay, have
a precise correspondence in Theory B, where the associated magnetically
charged dyon states decay. In the gauge theory, these
curves correspond to regions were the dimension of the monopole moduli
space enlarges discontinuously corresponding to the freedom for the
dyon to separate into its decay products \cite{tim,tim2}.   
Away from weak 
coupling, the determination of which states appear in the BPS spectra 
of either case is complicated by the presence of additional curves of marginal stability. 
Because the exact central charges agree, the same curves of marginal 
stability occur in both theories. This suggests, but does not guarantee, that 
the two spectra agree throughout the parameter space. In the next 
Section we will consider a brane configuration which provides further 
evidence for this proposal. 
 
One interesting consequence of the proposed equivalence is obtained by setting 
all twisted masses to zero. This leads to a novel 
connection between four-dimensional ${\cal N}=2$ SQCD with $N$ colours 
and $N_{f}=N+\tilde{N}$ massless flavours at the baryonic root 
and the supersymmetric 
$CP^{N-\tilde{N}}$ $\sigma$-model in two-dimensions. 
Both theories are asymptotically free and have a $U(1)_{R}$ symmetry 
which 
broken to a $Z_{2N-2\tilde{N}}$ subgroup by instantons. 
According to \cite{aps} 
the global symmetry on the baryonic branch is $SU(N-\tilde{N})$ which 
is the same as the global symmetry of the $CP^{N-\tilde{N}}$ 
$\sigma$-model. In addition to the 
massless spectrum of the four-dimensional theory described above, we 
should find massive hypermultiplet corresponding to each state in the 
$\sigma$-model spectrum. Thus we predict massive hypermultiplets 
transforming in the fundamental representation 
of the global symmetry group as well as bound states transforming 
in each antisymmetric tensor representation. In Appendix B, we show
that anti-symmetric tensor multiplets 
of the larger symmetry group $SU(N_f)$ exist at weak coupling for zero
masses but large VEVs, i.e.~{\it not\/} at the baryonic root.

\section{The Brane Configuration}

The technique of realising field theories as the world-volume 
dynamics of intersecting brane configurations 
in type II string theory \cite{hw} has 
proved very powerful in recent years, providing a geometrical 
description of many strong-coupling phenomena in field theory. 
In fact it is well known how to construct either Theory A or Theory B 
introduced above on the world-volume of 
type IIA branes which are subsequently elevated 
to M-theory. The four-dimensional 
theory was first studied in this context by Witten in \cite{witm5}, 
where it was shown that the M5-brane provides a 
concrete realisation of 
the Seiberg-Witten curves. The brane model of the two 
dimensional theory was studied by Hanany and Hori \cite{hh}. 
In the following 
sections we review both of these constructions and in, particular, how 
BPS spectrum is realized in each case. 
In both theories, BPS states correspond to M2 branes with boundaries 
on M5 branes. In both cases, the relevant boundaries are non-trivial 
one-cycles on the same Riemann surface.

We look first at the brane picture for the two-dimensional theory 
described in \cite{hh}. The configuration   
involves a pair of non-parallel NS 5-branes. The 
first of the pair spans world-volume directions $012345$ 
and is positioned at $x^6=x^7=x^8=x^9=0$. The second NS 5-brane 
is rotated with respect to the first, spanning world-volume directions  
$014589$. Such a rotation of the second NS 5-brane 
is typical when constructing theories with four supercharges and this 
brane is often referred to as an NS$^\prime$ 5-brane. It is located at  
$x^2=x^3=0$, and at some fixed value of $x^6$ and $x^7$. 
The configuration also involves $N+\tilde{N}$ semi-infinite D4-branes 
ending on the NS 5-brane.  
Each spans world-volume directions $01236$ and  
is located at $x^7=x^8=x^9=0$ and at a fixed value of $x^4$ and 
$x^5$. The D4-branes may end on the NS 5-brane either from the 
``left'' (world-volume $x^6<0$) or from the ``right'' ($x^6>0$). 
For our purposes we require $N$ to end from the right, and 
the remaining $\tilde{N}$ to end from the left. We will refer to 
the former as D4-branes and the latter as $\tilde{\rm D4}$-branes. 
A recent paper which discusses similar 
configurations of  semi-infinite D4-branes is \cite{tatar}. 

The $U(1)$ two-dimensional gauge theory is realised as the 
low-energy dynamics of a single D2-brane suspended between the NS 
and NS$^\prime$ 5-brane, with world-volume coordinates 017. 
The D2-brane has finite extent in the $x^7$ direction and 
is infinite in $x^0$ and $x^1$, these 
latter dimensions playing the role of the field theory space-time. 
The final configuration is depicted in figure 1. 
It preserves four supercharges as required for ${\cal N}=(2,2)$ 
supersymmetry in two-dimensions. 
Hanany and Hori argue that strings stretched between the D2-brane 
and semi-infinite D4-branes become chiral multiplets in two 
dimensions leading to Theory A described in the introduction.  

\begin{figure}
\begin{center}
\epsfxsize=4.0in\leavevmode\epsfbox{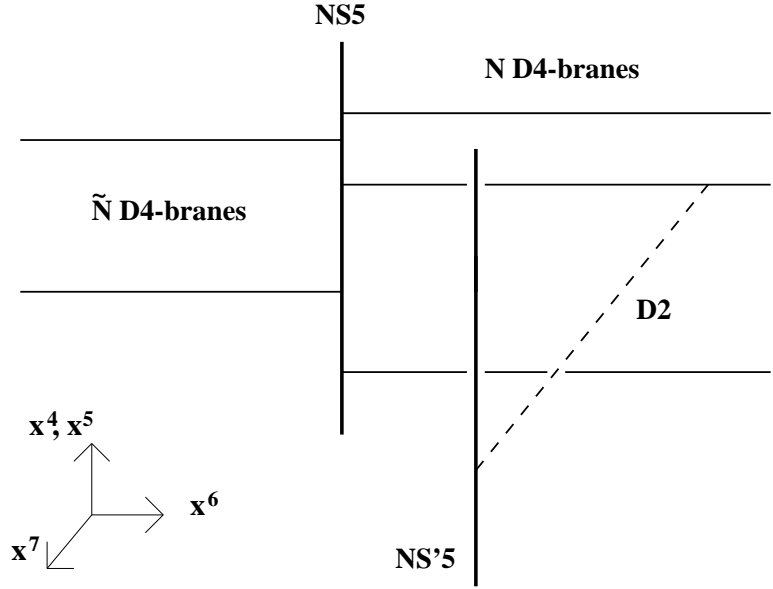}
\end{center}
\caption{The IIA configuration for theory A}
\elabel{IIA2}
\end{figure}
 
Parameters of the field theory are encoded as positions of 
the various branes. The VEV of the 
vector multiplet scalar and the twisted mass parameters 
both correspond to the positions of Dirichlet branes in the 
$(x^4,x^5)$ plane,
\be
\left. x^4+ix^5\right|_{\rm D2}&=&\sigma \nn\\ 
\left. x^4+ix^5\right|_{{\rm D4}_i}&=&m_i\ \ \ \ \ \ i=1,\ldots,N \nn\\
\left. x^4+ix^5\right|_{\tilde{D}4_{\tilde{i}}}&=&\tilde{m}_{\tilde{i}}
\ \ \ \ \ \ \tilde{i}=1,\ldots,\tilde{N}.
\nn\ee
Rotational symmetry in the $45$-plane is identified with the $U(1)_{A}$ 
R-symmetry of the ${\cal N}=(2,2)$ theory.  
The distances between the 
two 5-branes in the $x^6$ and $x^7$ directions are related to the 
bare FI parameter and bare gauge coupling respectively, 
\be
\frac{\Delta x^6}{g_{st}}&=&r \nn \\
\frac{\Delta x^7}{g_{st}}&=&\frac{1}{e^2}
\elabel{deltax}\ee
where $g_{st}$ is the string coupling constant, 
$\Delta x=x({\rm NS}^\prime)-x({\rm NS})$, and we are working in units 
where the string length-scale $\sqrt{\alpha'}$ is set to one. 
For $\Delta x^6=0$, both ends 
of the D2-brane lie on a 5-brane, and it is free roam the 
$(x^4,x^5)$ plane. This corresponds to the ``Coulomb branch'' of the 
gauge theory. For $\Delta x^6>0$, the D2-brane must end on one of 
the D4-branes and is therefore restricted to lie at one of $N$ 
fixed positions in $x^4$ and $x^5$ (this situation is shown in 
figure 1). 
For generic mass terms, these positions 
correspond to the $N$ classical vacua of the field 
theory for $r>0$ described in Section 2. Similarly, 
for $\Delta x^6<0$ there $\tilde{N}$ choices of $\tilde{\rm D4}$-brane 
on which the D2-brane may terminate. 
 
While the type IIA picture captures many of the 
classical properties of the field theory, including the 
classical moduli space of vacua, quantum effects 
are reproduced only after raising the configuration to M-theory 
\cite{hh}. 
As noted in that reference, the field theory limit of M-theory differs 
from the supergravity limit. Nevertheless we expect certain 
quantities, in particular masses of BPS states, to be correctly 
determined in the supergravity approximation. 
In eleven dimensions the system comprised of the NS 5-brane,  
D4-branes and $\tilde{D}4$-branes combines to become a single 
M5-brane, while the 
NS$^\prime$ 5-brane evolves into a flat M5-brane and the D2-brane becomes 
an M2-brane suspended between the two. 
We deal first with NS 5-brane and D4 ($\tilde{\rm D4}$)-branes. 
The world-volumes 
of all these branes share the $0123$ coordinates and the 
resulting M5-brane is correspondingly flat in these directions. 
The remaining world-volume directions of the M5-brane describe 
a two-dimensional curve, $\Sigma$, lying in the submanifold $R^3\times S^1$, 
parametrised by $x^4, x^5, x^6$ and periodic coordinate 
$x^{10}=x^{10}+2\pi R$. 
The preservation of eight supercharges requires  
this curve to be embedded holomorphically with respect 
to the complex coordinates
\be
v=x^4+ix^5\ \ \ ,\ \ \ s=R^{-3/2}x^6+iR^{-1}x^{10}
\elabel{complex}\ee
In terms of $t=\Lambda^{N}\exp(-s)$, the curve $\Sigma$ takes the form 
\cite{hh} 
\be
t\Lambda^{N-N_f}\prod_{\tilde{i}=1}^{\tilde{N}}(v-\tilde{m}_{\tilde{i}})
=\prod_{i=1}^{N}(v-m_i)
\elabel{mtheorya}\ee
where the product terms can be understood as 
arising from the deformation of the IIA NS 5-brane due to presence 
of the $D4-$ and $\tilde{\rm D4}$-branes. 
 
The description of the NS$^\prime$ 5-brane in M-theory is much 
simpler. It is not deformed by D4-branes and evolves 
to a flat M5-brane with world-volume directions 
$014589$. It's position in the $x^6$ and $x^{10}$ directions 
is given by
\be
t=\Lambda^{N}
\elabel{flatm5}\ee
Unlike the type IIA picture, the M5-branes no longer 
have a definite $x^6$ separation and the quantity $\Delta x^6(v)$ is 
interpreted as the running FI parameter in agreement with 
field theory \cite{hh}. In addition the curved M5 brane configuration 
is no longer invariant under rotations in the $45$-plane. This corresponds to 
the $U(1)_{A}$ anomaly which appears at one-loop in the two-dimensional 
quantum theory.    
 
The M-theory description of the D2-brane is as an M2-brane, 
stretched between the flat and curved M5-branes. Like its 
D2 descendant, the M2-brane world-volume is infinite in the $01$ 
directions and finite in $x^7$. Unlike the M5-branes however, the 
M2-brane does not have a fixed configuration. Rather, different 
configurations correspond to different states of the 
two-dimensional gauge theory. For example, supersymmetry is 
unbroken in the vacuum state of the gauge theory and this is 
reflected in the corresponding 
M2-brane ground configuration by the requirement that it 
preserve four supercharges. The only such M2-brane   
configurations are straight strips lying at a fixed value of  
$v$ and $t$. It must intersect the two M5-branes and 
therefore has to lie at $t=\Lambda^{N}$. The possible positions 
in the $v$-plane are then determined by
\be 
\prod_{i=1}^{N}(v-m_i)-\Lambda^{N-\tilde{N}}
\prod_{\tilde{i}=1}^{\tilde{N}}(v-\tilde{m}_{\tilde{i}})=0
\elabel{vac}\ee
in agreement with field theoretic expectations \eqn{vacuumeq3}. 
Excited states in the field theory correspond to excited states 
of the M2-brane, no longer a straight strip but a topologically 
non-trivial surface with boundaries on the two M5-branes. 
We will 
return to these states after first reviewing the 
four-dimensional theory. 
 
The type IIA realisation of pure $SU(N)$ ${\cal N}=2$ SYM consists 
of $N$ D4-branes suspended between two parallel NS 5-branes. 
The latter span world-volume directions $012345$. 
Each is located at $x^7=x^8=x^9=0$ and at a fixed $x^6$ value. 
The separation of the two NS5-branes in the $x^6$ direction 
determines the bare four-dimensional coupling constant,
\be
\frac{\Delta x^6}{g_{st}}=\frac{1}{g^2}
\elabel{g2x6}\ee
The four-dimensional 
theory itself lives on the world-volume of the D4-branes, spanning 
directions $01236$. 
The D4-branes are finite in the $x^6$ direction and at large 
distances their low-energy dynamics are described by $D=4$, $SU(N)$ 
${\cal N}=2$ SYM with $x^0,\ldots,x^3$ playing the role 
of space-time \cite{witm5}. The D4-branes are free to move in the 
$v$-plane and their positions in these directions parametrise the 
classical Coulomb branch of the theory.
 
Hypermultiplets transforming in the fundamental representation 
of the gauge group may be incorporated in one of two ways: using 
either D6-branes or D4-branes. We choose the latter. The world-volumes 
of these D4-branes also span directions $01236$. Each is attached to 
one of the two NS 5-branes and are thus semi-infinite in 
the $x^6$ direction. In order to exhibit the similarity to theory 
A in the most transparent manner, we choose to 
attach $N$ semi-infinite D4-branes to the ``right-hand'' NS 5-brane. 
These have world-volume $x^6\rightarrow+\infty$.  
The remaining $\tilde{N}$ D4-branes are attached to the ``left-hand'' 
NS 5-brane, have world-volume $x^6\rightarrow -\infty$ and, in 
analogy with theory A, will be referred to as $\tilde{\rm D4}$-branes. 
The position of each semi-infinite D4-brane
in the $v$-plane determines the mass of the corresponding 
hypermultiplet.
\be
\left. x^4+ix^5\right|_{{\rm D4}_i}&=&m_i\ \ \ \ \ \ i=1,\ldots,N. \nn\\
\left. x^4+ix^5\right|_{\tilde{\rm D4}_{\tilde{i}}}&=&
\tilde{m}_{\tilde{i}}\ \ \ \ \ \ i=1,\ldots,N_f-N.
\nn\ee
As explained in the previous section, the root of the baryonic 
Higgs branch occurs classically when the 
VEVs are equal to the masses of some set of $N$ hypermultiplets. 
The particular brane configuration described above above 
naturally selects $N$ hypermultiplets as those ending on the 
``right-hand'' NS 5-brane. The root of this baryonic Higgs branch 
then corresponds to the situation depicted in figure 2.

\begin{figure}
\begin{center}
\epsfxsize=5.0in\leavevmode\epsfbox{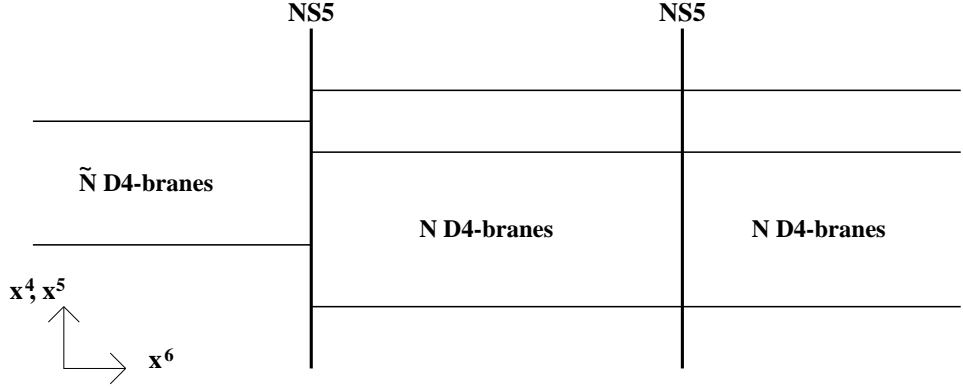}
\end{center}
\caption{The IIA configuration for theory B}
\elabel{IIA}
\end{figure}

As in the two-dimensional case, many quantum aspects of the 
field theory are uncovered by elevating this configuration to 
eleven dimensions. Generically, such a system of NS 5-branes 
and D4-branes becomes a single smooth object in M-theory. The 
world-volume of this single M5-brane spans directions $0123$ and 
describes a two-dimensional surface embedded in $x^4, x^5, x^6$ 
and $x^{10}$. Once again, the preservation of eight supercharges 
requires the curve to be holomorphic with respect to the 
complex structures \eqn{complex}.
 
For the IIA configuration corresponding to an arbitrary point 
on the Coulomb branch, the lift to M-theory is described by
\be
F(t,v)=A(v)t^2+B(v)t+C(v)=0
\elabel{Ftv}\ee
where
\be
A(v)&=&\Lambda^{N-N_f}\prod_{\tilde{i}=1}^{N_f-N}
(v-\tilde{m}_{\tilde i}) \nn\\
B(v)&=&\prod_{a=1}^{N}(v-\phi_a) \nn\\
C(v)&=&\Lambda^{N}\prod_{i=1}^{N}(v-m_i) \nn
\ee
We will consider the asymptotically free 
case, $N_f<2N$. The curve branches at $B^2-4AC=0$, defining the 
$2N$ branch points, $e_a$ and $\tilde{e}_a$,
\be
\prod_{a=1}^{N}(v-\phi_a)^2-4\Lambda^{2N-N_f}\prod_{i=1}^{N}
(v-m_i)\prod_{\tilde{i}=1}^{N_f-N}(v-\tilde{m}_{\tilde{i}})
\equiv\prod_{a=1}^{N}(v-e_a)(v-\tilde{e}_a)
\elabel{one}\ee
The root of the baryonic Higgs branch occurs at the point at which a  
maximal number of cycles vanish. This occurs when $e_a=\tilde{e}_a$ 
for all $a=1,\ldots,N$. The solution is given by
\be 
B(v)=-\Lambda^{2N-N_f}\prod_{\tilde{i}=1}^{N_f-N}
(v-\tilde{m}_{\tilde{i}})-\prod_{i=1}^{N}(v-m_i)
\nn\ee
and the curve becomes,
\be
F(t,v)=\left(t\Lambda^{N-N_f}\prod_{\tilde{i}=1}^{N_f-N}
(v-\tilde{m}_{\tilde{i}})
-\prod_{i=1}^{N}(v-m_i)\right)\left(t-\Lambda^{N}\right)
\elabel{Ftv3}
\ee
This is the curve used in Section 4 to obtain the exact BPS spectrum 
of Theory B. Importantly, at the root of the baryonic Higgs branch the curve 
has a very simple factorisation. 
The single M5-brane can be thought of as degenerating 
into two M5-branes, each described by one of the factors of \eqn{Ftv3}. 
The first factor corresponds to a non-trivial M5-brane, while the 
second is simply a flat M5-brane located at $t=\Lambda^{N}$. 
Notice that 
the first factor is precisely the curve $\Sigma$ describing the 
NS 5-brane and 
D4-branes of the two-dimensional theory \eqn{mtheorya}. Of course, 
this is not surprising considering the similarity of the type IIA 
configurations for theories A and B. 
\paragraph{}
We now turn to the description of BPS states in the two theories, starting 
with Theory A \cite{hh}. 
States of the field theory correspond to states of the M2-brane. 
We have already seen how the ground states of the field 
theory are reproduced by the M2-brane states \eqn{vac} preserving four 
supercharges. The M-theory realisation of BPS states may be 
identified in a similar manner as M2-brane configurations 
preserving two supercharges. In Appendix C we 
show that this condition requires the spatial part of the 
M2-brane world-volume form to be the pull-back of 
\be
\omega=\Omega+{\rm d}x^7\wedge{\rm d}x^1
\elabel{oneform}\ee
where $\Omega$ is the holomorphic two-form
\be
\Omega={\rm d}s\wedge{\rm d}v
\elabel{omega}\ee
The mass of such a BPS state is given by the difference between the 
mass of the corresponding excited M2-brane and the ground M2-brane.
Each is given by the area of the brane multiplied by the brane tension, 
$T$, and may be regulated by introducing a cut-off on the 
integration over $x^1$. The ground state brane, 
which we denote as M2$_0$, is at fixed values of both $v$ and $t$ 
and has world-volume given by the pull-back of 
${\rm d}x^7\wedge{\rm d}x^1$. The mass of any BPS state 
is therefore given by
\be
M&=&T\left|\int_{\rm M2}\omega\ - \int_{{\rm M2}_0} 
\omega\ \right| \nn\\
&=&T\left|\int_{\rm M2}\Omega\ + \int_{{\rm M2}-{\rm M2}_0}
{\rm d}x^1\wedge{\rm d}x^7\ \right| \nn\\
&=&T\left|\int_{\rm M2}\Omega\ \right|
\nn\ee
This formula for the mass of BPS states was also proposed in 
\cite{hh} using different arguments. The mass is seen to 
be independent of $\Delta x^7$ in agreement with field theory 
expectations that the masses of BPS states do not depend on the 
gauge coupling constant, $e$.  To make contact with the 
BPS mass formula of Section 3, note that $\Omega$ 
is exact; $\Omega= {\rm d}\lambda_{SW}$, where
\be
\lambda_{SW}=v\frac{{\rm d}t}{t}
\elabel{lambda}\ee
is recognised, in the four-dimensional context, as the 
Seiberg-Witten differential introduced in Section 4. 
Using Stokes' theorem, the mass of the BPS states can be rewritten 
as 
\be
M=T\left|\int_{\partial {\rm M2}}\lambda_{SW}\ \right|
\elabel{masscycle}\ee
where $\partial$M2 is the boundary of the M2-brane. For 
$\Delta x^7=0$, this boundary consists of two curves, 
each lying on one of the M5-branes. Denote the curve on the 
flat M5-brane as $C_r$ and that on the curved M5-brane as $C_l$. Then,  
\be
M=T\left|\int_{C_l+C_r}\lambda_{SW}\ \right|
=T\left|\int_{C_l}\lambda_{SW}\ \right|
\nn\ee
where the second equality follows 
because $\lambda_{SW}$ vanishes over the flat M5-brane \eqn{flatm5} 
positioned at a fixed value of $t$. Note that the masses of the BPS states 
depending only on the details of the curved M5 brane $\Sigma$ 
\eqn{mtheorya}. 
The M-theory realisation of  BPS states in the four-dimensional 
theory were discussed in detail 
in \cite{henyi,mik}. They also correspond to holomorphically embedded 
M2-branes with world-volume form given by the pull-back of 
$\Omega$ and with boundary on the M5-branes. The mass of 
such a state is given by the integral of the Seiberg-Witten 
differential over the boundary of the M2-brane, in 
agreement with \eqn{masscycle}. Moreover, the factorisation of the 
Seiberg-Witten curve at the baryonic root ensures that, as in Theory 
A, the masses of BPS states are determined solely by a choice of 
boundary $C_l$ on $\Sigma$, the curved part of the M5-brane \eqn{Ftv3}.
\paragraph{}
Although we have shown both in field theory and from branes that the 
BPS mass formula for the two theories agrees, we would like conclude 
that the quantum numbers of BPS states appearing in the two theories 
are the same throughout parameter space. As mentioned in the previous 
sections, there is no obvious way to do this using 
conventional field theory methods. In the brane context, determining 
which BPS states appear is equivalent to determining which BPS configurations 
of the M2 brane are allowed. In the present case this amounts to determining 
which one-cycles on the curve $\Sigma$ can arise as boundaries of a BPS 
M2 brane. At least for small masses, this problem was solved 
explicitly for Theory A in \cite{hh}, yielding agreement with previous 
results for the $CP^{N-1}$ $\sigma$-model. Here we simply note that 
the equivalent problem for Theory B is essentially identical and the 
same analysis should apply. This provides further evidence for the exact 
agreement of the BPS spectra of the two theories.    
\paragraph{}
Finally, we should comment on the fact that not all BPS states 
of Theory B appear in spectrum of Theory A. Specifically, 
only the massive hypermultiplets of Theory B participate in the 
correspondence. The fact that the massless states of 
Theory B have no counterparts in Theory A corresponds to a general feature 
of configurations involving branes of different dimensions which was first 
discussed in \cite{hw}: moduli in the higher-dimensional world-volume theory 
correspond to parameters in the lower-dimensional world-volume theory. 
By ${\cal N}=2$ SUSY all massless fields in Theory B 
are associated with scalar moduli. As these moduli become parameters 
in Theory B, we know that the corresponding massless 
degrees of freedom must decouple from the two-dimensional spectrum. 
The question of what happens to the massive vector multiplets of Theory 
B is more subtle as these states have exactly the same masses and charges 
as some of the hypermultiplets at the baryonic root. Because of this 
degeneracy, it seems possible that a 
massive vector multiplet can be interpreted as a threshold 
boundstate of a massive hypermultiplet and some massless ones. 
In principle, the existence of such threshold bound-states 
may be sensitive to deformations of the theory, such as those 
considered above, which leave the spectrum of truly 
stable BPS states invariant. Clearly this aspect of the correspondence 
requires further investigation.

\subsection*{Acknowledgements}

The authors would like to thank Sumit Das, 
Sunil Mukhi, Sumati Surya, Brett Taylor, Sachin Vaidya and 
especially Ami Hanany for useful discussions. 
We would also like to thank Radu Tatar for providing us with a 
preliminary version of \cite{tatar}. ND and TJH acknowledge support from 
the TMR network grant FMRX-CT96-0012.

\section*{Appendix A}
\renewcommand{\theequation}{A.\arabic{equation}}
\setcounter{equation}{0}
 
In this appendix we give some details of the soliton solutions
of the theory in terms of the $({\cal R},\Theta)$ variables of equation 
(\ref{N=2lag}). From (\ref{bog1}) we deduce that a BPS saturated 
soliton solution satisfies the first order equation, 
\begin{equation}
\frac{\partial\Lambda}{\partial x }=
\frac{\Delta{\cal W}}{|\Delta{\cal W}|} g^{\Lambda\bar{\Lambda}}
\frac{\partial\bar{{\cal W}}}
{\partial\Lambda^{\dagger}}=\exp(i\gamma)\frac{|m|}{r}{\cal R}({\cal R}-r)
\elabel{bogeqN=2}
\end{equation}
where, $\gamma=\tan^{-1}(2\pi r/\Delta\Theta)$. 
Taking the real and imaginary parts of this equation we obtain, 
\begin{equation}
\frac{\partial {\cal R}}{\partial x}=\frac{1}{2\pi}\tan\gamma\frac
{\partial \Theta}{\partial x}
=|m|\sin\gamma\left[\frac{{\cal R}(r-{\cal R})}{r}\right]
\elabel{eom2}
\end{equation}
The required solution obeys the boundary conditions, 
$\Lambda\rightarrow 0$ as $x\rightarrow -\infty$ and 
$\Lambda\rightarrow \Lambda_{+}=ir+\Delta\Theta/2\pi$ 
as $x\rightarrow +\infty$ and is given by, 
\begin{eqnarray}
{\cal R}  = \frac{r\exp(|m|\sin\gamma x)}{1+\exp(|m|\sin\gamma x)} 
 &\qquad{} \qquad{} & 
\Theta=\Delta \Theta \frac{\exp(|m|\sin\gamma x)}{1+\exp(|m|\sin\gamma x)}
\elabel{rtsoln}
\end{eqnarray}
which describes a straight line-segment in the $\Lambda$ plane which joins the points $\Lambda=0$ and $\Lambda=\Lambda_{+}$. The resulting central charge 
is just 
$Z=2\Delta{\cal W}=im\Delta\Lambda=im(i\Delta{\cal R}+\Delta\Theta/2\pi)$
and BPS mass formula implies 
that the mass of the soliton is equal to the length of this line-segment   
\begin{equation}
M=|Z|=|m|\sqrt{r^{2}+ \left(\frac{\Delta \Theta}{2\pi}\right)^{2}}
\elabel{mass3}
\end{equation}
where we have used $\Delta{\cal R}=r$. 
To interpret the meaning of quantity $\Delta \Theta$ we note that, in 
the $(\Theta,\alpha)$ variables of (\ref{sg}), the Noether charge, $S$, 
of the global $U(1)$ symmetry of the theory is given as,
\begin{equation}
S=\frac{r}{2}\int_{-\infty}^{+\infty}dx\,\sin^{2}\varphi\,\dot{\alpha} + 
\frac{\theta}{4\pi}\int_{-\infty}^{+\infty}dx\, \partial_{x}(\cos\varphi)
\elabel{scharge}
\end{equation}
On the other hand from (\ref{rtsoln}) and (\ref{ident}) we have,   
\begin{equation}
\frac{1}{2\pi}\frac{\partial \Theta}{\partial x}=\frac{r}{2}\sin^{2}
\frac{\varphi}{2}\frac{\partial \alpha}{\partial t}
\end{equation}
on integrating this equation we find, 
\begin{equation}
\frac{\Delta \Theta}{2\pi}=\frac{r}{2}\int_{-\infty}^{\infty}dx\,
\sin^{2}\varphi\frac{\varphi}{2}\frac{\partial \alpha}{\partial t}=
S+\frac{\theta}{2\pi}
\end{equation}
Thus we recover the central charge (\ref{ccharge4}) and the 
mass formula (\ref{sat2}), 
\begin{equation}
M=|m|\sqrt{ \left(S+\frac{\theta}{2\pi}\right)^{2}+ r^{2}}
\elabel{mass4}
\end{equation}

\section*{Appendix B}
\renewcommand{\theequation}{B.\arabic{equation}}
\setcounter{equation}{0}
 
In this appendix, we construct the weak coupling spectrum of
bound-states of dyons and quarks in an ${\cal N}=2$ supersymmetric 
$SU(N)$ gauge theory with $N_f$ hypermultiplets. Consider the monopole
solution with topological (magnetic) charge $h_a=\delta_{ac}-\delta_{ab}$. At a
generic point in the space of Higgs
VEVs, the moduli space of such solutions will be
completely described by embedding an $SU(2)$ monopole solution in the
$SU(N)$ gauge group \cite{tim}. In our analysis, we shall avoid submanifolds of
co-dimension one, where the space of solutions enlarges
discontinuously. These submanifolds correspond to regions where some
$\phi_d$ lies on the line segment joining $\phi_b$ with $\phi_c$ and
are surfaces on which dyons can decay to other dyons.

Before we proceed, we find it useful to introduce the roots and
weights of $SU(N)$. To this end, we introduce $\Be_a$, $a=1,\ldots,N$,
the weights of the $\BN$ representation. The roots are then 
$\Be_a-\Be_b$. The
monopole that we are considering has a topological charge equal to the
root $\Bh\equiv\Balpha=\Be_c-\Be_b$.
We now analyse the time-independent Dirac equation for the fermion field
$\Psi^\lambda_a$, in the hypermultiplet $\{Q_a^\lambda,\tilde
Q^a_\lambda\}$, in the background of the monopole. The resulting equation
is only non-trivial if $a=b$ or $a=c$ and without loss of generality
we can choose $a=b$. The Dirac equation
is given by a generalization of that written down in \cite{tim} which
includes an explicit mass term:
\begin{equation}
\begin{pmatrix} S & -{\cal D}^* \\ -{\cal D} & -S\end{pmatrix}
\Psi_b^\lambda=E\Psi_b^\lambda
\elabel{direq}\end{equation}
Here 
\begin{equation}
{\cal D}=-i\vec\sigma\cdot\vec D+iP\ ,\qquad D^*=-i
\vec\sigma\cdot\vec D-iP
\elabel{derivs}\end{equation}
where $\vec\sigma$ are the Pauli sigma matrices and $\vec D$ is the
spatial part of the covariant derivative in the background of the
monopole solution. The other quantities in (\ref{direq}) and
(\ref{derivs}) are
\begin{eqnarray}
P&=&\phi^ct^c+{\rm
Re}(e^{-i\sigma}\Bphi)\cdot\BH-\frac{|\Bphi\cdot\Balpha|}{\Balpha^2}\Balpha\cdot\BH+{\rm
Re}(e^{-i\sigma}m_\lambda)\nonumber\\
S&=&{\rm Im}(e^{-i\sigma}\Bphi)\cdot\BH+{\rm Im}(e^{-i\sigma}m_\lambda)
\end{eqnarray}
where $\sigma$ is the phase of $\Bphi\cdot\Balpha$, 
the $t^c$ are the generators of the $SU(2)$ corresponding to the
root $\Balpha$, $\phi^c$ is the Higgs field of the $SU(2)$ BPS
monopole and $\BH$ are the Cartan generators of $SU(N)$.

We will now show that BPS bound-states of the monopole, and its dyonic
excitations, with the quark
exists whenever ${\cal D}$ has a normalizable zero-mode. Suppose that
such a mode $\psi_b^\lambda$ exists. This implies that the Dirac operator has an
eigenvector:
\begin{equation}
\begin{pmatrix} S & -{\cal D}^* \\ -{\cal D} & -S\end{pmatrix}\left({
\psi_b^\lambda\atop 0}\right)={\rm Im}\left(e^{-\sigma}\Bphi\cdot\Be_b+e^{-\sigma}m_\lambda\right)
\left({\psi_b^\lambda\atop 0}\right)
\end{equation}
where we used the fact that $\BH\Psi^\lambda_b=\Be_b\Psi^\lambda_b$. To
show that the bound-state is BPS saturated, we notice that although the mode is not
a zero-mode of the full Dirac operator, but only of ${\cal D}$, it
contributes to the central
charge in just such a way as to preserve the BPS condition. The central charge of the bound-state
is $Z=\tau\Bphi\cdot\Balpha+\Bphi\cdot\Be_a+m_\lambda$ and hence the
BPS mass is
\begin{equation}
M=\big|\tau\Bphi\cdot\Balpha+\Bphi\cdot\Be_b+m_\lambda\big|
=\big|\tau\Bphi\cdot\Balpha\big|+{\rm
Im}\left(e^{-i\sigma}\Bphi\cdot\Be_b+e^{-i\sigma}m_\lambda\right)+\cdots
\end{equation}
Here the first term is the mass of the monopole, the second term is
precisely the energy of the fermion mode and the ellipsis are
corrections in $g$ which encode the back-reaction of the fermion
fields on the monopole that we have not accounted for in our simple analysis.
Notice the this line of reasoning is analogous to our analysis of the
soliton-fermion bound-states in Theory A. The semi-classical description of the
bound-state is completely standard (see for example
\cite{harvey}). The fermion mode is associated to a pair of  raising
and lowering operators $\rho$ and $\rho^\dagger$ with canonical
anti-commutation relation $\{\rho,\rho^\dagger\}=1$. The monopole
carries a Fock space representation of these operators, in other words
there is a vacuum state $|0\rangle$, which represents the original
monopole, and an excited state $\rho^\dagger|0\rangle$, which
represents the bound-state. In addition, the monopole wavefunction
also includes the usual dyonic excitation piece which implies that
each bound-state gives rise to a tower of states with magnetic charge 
$\Bh=\alpha$, electric charge $\Bq=S\Balpha+\Be_b$, for an integer $S$,
and global charge $s_\mu=\delta_{\mu\lambda}$.

So our problem is now reduced to searching for normalizable zero-modes
of ${\cal D}$. Fortunately this analysis is a simple generalization of
that in \cite{tim}. There exists a single normalizable zero-mode if
\begin{equation}
-1<{\rm Re}\left(\frac{\Bphi\cdot(\Be_b+\Be_c)+2m_\lambda}{\Bphi\cdot\Balpha}\right)<1
\elabel{cond}\end{equation}

We now analyse this condition at the baryonic root, where
$\phi_a\equiv\Bphi\cdot\Be_a=-m_i\delta_{ia}$. Furthermore, in order
to facilitate comparison with Theory A, we shall write the topological charge
of the dyon as $\Balpha=\Be_k-\Be_l$, and so
$\Bphi\cdot\Balpha=m_l-m_k$, and the fermion to have
mass $m_\lambda$ equal to (i) $m_j$ (ii) $\tilde m_{\tilde j}$. For
the first case the region where the bound-state exists is
\begin{equation}
0<{\rm Re}\left(\frac{m_l-m_j}{m_l-m_k}\right)<1
\label{reg5}\end{equation}
with quantum numbers $T_i=\delta_{il}-\delta_{ik}$,
$S_i=ST_i+\delta_{il}-\delta_{ij}$ and $\tilde S_{\tilde i}=0$. While in the second case 
the bound-state exists in a region
\begin{equation}
0<{\rm Re}\left(\frac{m_l-\tilde m_{\tilde j}}{m_l-m_k}\right)<1
\label{reg6}\end{equation}
with quantum numbers $T_i=\delta_{il}-\delta_{ik}$,
$S_i=ST_i+\delta_{il}$ and $\tilde S_{\tilde i}=-\delta_{\tilde
i\tilde j}$.
In overlapping regions (\ref{reg5}) and (\ref{reg6}) there will be
multiple bound-states obtaining by filling out the states of a higher
dimensional Fock space.

It is also interesting to analysis the spectrum of bound-state with
zero masses $m_\lambda=0$. In this case the condition for the
existence of a bound-state is
\begin{equation}
0<{\rm Re}\left(\frac{\phi_l}{\phi_l-\phi_k}\right)<1
\label{reg7}\end{equation}
In this region, there are $N_f$ degenerate fermion modes
$\psi_l^\lambda$, $\lambda=1,\ldots,N_f$, and the bound-states are
composed of a Fock space representation of the $N_f$ creation and
annihilation operator $\rho_\lambda$ and $\rho_\lambda^\dagger$. The
states are therefore of the form
\begin{equation}
\rho_{\lambda_1}^\dagger\cdots\rho_{\lambda_p}^\dagger|0\rangle\qquad
\lambda_i\neq\lambda_j
\end{equation}
For a given $p$, these states transform in the
$p^{\rm th}$ anti-symmetric representation of the unbroken $SU(N_f)$
flavour symmetry group.

\section*{Appendix C}
\renewcommand{\theequation}{C.\arabic{equation}}
\setcounter{equation}{0}
   
In this appendix, using the methods of \cite{bbs}, we 
derive the world-volume form \eqn{oneform} 
for a M2-brane preserving two supercharges. 
We follow closely the analogous calculation for 
BPS states in four-dimensional ${\cal N}=2$ theories performed 
by Fayyazuddin and Spali\'nski \cite{fs}. A convenient choice 
of eleven dimensional gamma matrices $\Gamma_M$ satisfying 
$\{\Gamma_M,\Gamma_N\}=2\delta_{MN}$ 
is given in the appendix of this reference, such that $\Gamma_M$ 
are real for $M=2,5,7,8,10$ and purely imaginary 
for other space-time indices and all gamma matrices with spatial 
indices are hermitian.   
 
The number of supersymmetries preserved by an p-brane with embedding  
$X^M$ in $R^{1,9}\times S^1$ is equal to the number of 
solutions to
\be
\eta = \frac{1}{p!}\epsilon^{\alpha_1...\alpha_p}
\Gamma_{M_1...M_p}\partial_{\alpha_1}X^{M_1}...\partial_{\alpha_p}
X^{M_p}
\elabel{susy}\ee
where $\Gamma_{M_1...M_p}=\Gamma_{[M_1}...\Gamma_{M_p]}$ and 
$\eta$ is an eleven dimensional Majorana spinor (64 real 
components) which can be decomposed as 
\be
\eta=\chi+B\chi^\star
\nn\ee
where $B$ is equal to the product of the real gamma matrices and 
determines the charge conjugation matrix $C=B\Gamma_0$. 
We first consider solutions to \eqn{susy} in the background  
of the M5-branes. The NS 5-brane and D4-brane system lifts to 
the single  M5-brane described by the curve \eqn{mtheorya}. 
It was shown explicitly in \cite{fs} that this preserves eight 
supercharges satisfying 
\be
\Gamma_{\bar{v}}\chi=\Gamma_{\bar{s}}\chi&=&0 \nn\\
i\Gamma_0\Gamma_1\Gamma_2\Gamma_3\chi&=&-\chi
\elabel{first}\ee
where $\Gamma_{\bar{v}}$ and $\Gamma_{\bar{s}}$ are the gamma 
matrices in the complex basis \eqn{complex}. The flat M5-brane 
enforces a further condition on the spinor:  
$\Gamma_0\Gamma_1\Gamma_4\Gamma_5\Gamma_8\Gamma_9\eta=\eta$. 
Using the first equation in \eqn{first}, it is simple to show that 
this is equivalent to 
\be
i\Gamma_0\Gamma_1\Gamma_8\Gamma_9\chi=-\chi
\elabel{second}\ee
There are four real independent solutions to \eqn{first} and 
\eqn{second}, equivalent to ${\cal N}=(2,2)$ supersymmetry in 
two dimensions. Further, using the fact that the product of all 
gamma matrices is proportional to the identity matrix, any 
spinor $\chi$ satisfying both these conditions must also satisfy
\be
\Gamma_0\Gamma_1\Gamma_7\chi=\chi
\nn\ee
This reflects the fact that we may add a flat M2-brane to the 
configuration while preserving four supercharges. This M2-brane 
corresponds to the ground state of the two-dimensional field theory.
 
Each state in the field theory other than the vacuum corresponds 
to an excited M2-brane which is no longer embedded at a fixed value 
of $v$ and $s$. The 
condition that this state be BPS requires the existence of a 
simultaneous solution to \eqn{first}, \eqn{second} and
\be
\eta=\ft12\epsilon_{\alpha\beta}\Gamma_0\Gamma_{MN}\partial_\alpha
X^M\partial_\beta X^N\eta
\nn\ee
Introducing the complex coordinate $u=x^1+ix^7$, this condition 
becomes 
\be
\epsilon_{\alpha\beta}\eta&=&\Gamma_0\big\{
(\partial_\alpha s\partial_\beta v-\partial_\alpha v \partial_\beta 
s)\Gamma_{sv} + (\partial_\alpha \bar{s}\partial_\beta\bar{v}-
\partial_\alpha\bar{v} \partial_\beta\bar{s})\Gamma_{\bar{s}\bar{v}} 
\nn\\ && \ +\ 
(\partial_\alpha s\partial_\beta u-\partial_\alpha u \partial_\beta 
s)\Gamma_{su} + (\partial_\alpha \bar{s}\partial_\beta\bar{u}-
\partial_\alpha\bar{u} \partial_\beta\bar{s})\Gamma_{\bar{s}\bar{u}}
\nn\\ && \ +\   
(\partial_\alpha v\partial_\beta u-\partial_\alpha u \partial_\beta 
v)\Gamma_{sv} + (\partial_\alpha \bar{v}\partial_\beta\bar{u}-
\partial_\alpha\bar{u} \partial_\beta\bar{v})\Gamma_{\bar{v}\bar{u}} 
\nn\\ && \ +\  
(\partial_\alpha s\partial_\beta\bar{u}-\partial_\alpha\bar{u} 
\partial_\beta s)\Gamma_{s\bar{u}} + (\partial_\alpha \bar{s}
\partial_\beta u -\partial_\alpha u \partial_\beta\bar{s})
\Gamma_{\bar{s}u} 
\elabel{hmm}\\ && \ +\   
(\partial_\alpha v\partial_\beta\bar{u}-\partial_\alpha\bar{u} 
\partial_\beta v)\Gamma_{v\bar{u}} + (\partial_\alpha \bar{v}
\partial_\beta u -\partial_\alpha u \partial_\beta\bar{v})
\Gamma_{\bar{v}u} 
\nn\\ && \ + \  
i(\partial_\alpha s\partial_\beta\bar{s}+\partial_\alpha v 
\partial_\beta \bar{v} - \partial_\alpha\bar{s}\partial_\beta s 
-\partial_\alpha \bar{v}\partial_\beta v)\Gamma_0\Gamma_1\Gamma_2
\Gamma_3 
\nn\\ && \ + \ 
(\partial_\alpha u\partial_\beta\bar{u}-\partial_\alpha \bar{u}
\partial_\beta u)\Gamma_{u\bar{u}}\,\big\}\eta
\nn\ee
Mercifully many of these terms vanish upon applying the projection 
operator $Q=\Gamma_s\Gamma_{\bar s}\Gamma_v\Gamma_{\bar v}$, and 
we are left with
\be
\epsilon_{\alpha\beta}B\chi^\star &=& (\partial_\alpha s\partial_\beta v
-\partial_\alpha v \partial_\beta s)\Gamma_0\Gamma_s\Gamma_v\chi 
-i(\partial_\alpha u\partial_\beta\bar{u}-\partial_\alpha \bar{u}
\partial_\beta u)B\chi^\star 
\nn\\
&& + i(\partial_\alpha s\partial_\beta\bar{s}+\partial_\alpha v 
\partial_\beta \bar{v} - \partial_\alpha\bar{s}\partial_\beta s 
-\partial_\alpha \bar{v}\partial_\beta v)\Gamma_0B\chi^\star 
\nn\ee
Without the second term on the right-hand side, this is the 
equation  
determining the M2-brane realisation of a BPS state in the 
four-dimensional ${\cal N}=2$ theory. In the two-dimensional model 
under consideration, the presence of this second term reflects the 
spatial extension of the M2-brane in the $x^1$ and $x^7$ directions.
Notice that any term appearing in \eqn{hmm} that contained  
both $u$ and either $s$ or $v$ is eliminated by the projection. 
Finally, we act with a further projection, 
$P=\ft12 (1+i\Gamma_0\Gamma_1\Gamma_2\Gamma_3)$, to arrive at the 
requirement
\be
B\chi^\star=\Gamma_0\Gamma_s\Gamma_v\chi
\nn\ee
which is identical to the four-dimensional BPS condition \cite{fs}. 
Moreover, the world-volume form on the BPS M2-brane is determined to 
be
\be
\epsilon_{\alpha\beta}=\partial_\alpha s\partial_\beta v
-\partial_\alpha v\partial_\beta s -i\partial_\alpha u\partial_\beta
\bar{u}+i\partial_\alpha\bar{u}\partial_\beta u
\nn\ee
which is indeed the pull-back of the two form $\omega$ given in 
equation \eqn{oneform}.

\end{document}